\documentclass[12pt]{article}

\pdfoutput=1
\usepackage{amssymb}
\usepackage{amsmath}
\usepackage{latexsym}
\usepackage[usenames]{color}
\usepackage{cite}
\usepackage{setspace}
\definecolor{darkgreen}{rgb}{0,0.5,0}
\definecolor{darkblue}{cmyk}{0.9,0.9,0,0}
\definecolor{darkred}{rgb}{0.6,0,0.3}

\usepackage{graphicx} 
\usepackage{epsfig}
\usepackage[setpagesize=false,pagebackref=false, linktocpage, bookmarksopen=true, colorlinks=true, linkcolor=darkblue,citecolor=darkblue,urlcolor=darkblue]{hyperref}
\usepackage{hyperref}
\newcommand{\arXiv}[2]{\href{http://arxiv.org/abs/#1}{{\tt arXiv:#2}}}
\newcommand{\hep}[2]{\href{http://arxiv.org/abs/#1}{{\tt #2}}}

\newcommand{\tr}{{\rm tr}}

\renewcommand{\thefootnote}{\arabic{footnote}}

\def\del{\partial}

\def\fn#1{\footnote{#1}}
\def\nn{\nonumber}
\def\eqref#1{(\ref{#1})}
\def\comma{\,,}
\def\period{\,.}
\def\beq{\begin{equation}}
\def\eeq{\end{equation}}
\def\pmatrix#1#2{\left( 
\begin{array}{#1}
#2\end{array} 
\right)}

\DeclareSymbolFont{extraup}{U}{zavm}{m}{n}
\DeclareMathSymbol{\varheart}{\mathalpha}{extraup}{86}
\DeclareMathSymbol{\vardiamond}{\mathalpha}{extraup}{87}
\begin{document}
\thispagestyle{empty}

\renewcommand{\thefootnote}{\fnsymbol{footnote}}
\setcounter{page}{1}
\setcounter{footnote}{0}
\setcounter{figure}{0}
\begin{flushright}
{\sf UT-Komaba/17-6}
\end{flushright}
\begin{center}
$$$$
{\Large\textbf{\mathversion{bold}
Structure Constants of Defect Changing Operators on the $1/2$ BPS Wilson Loop
}\par}

\vspace{1.3cm}

\textrm{Minkyoo Kim$^{\textcolor[rgb]{0.9,0,0}{\varheart}}$, Naoki Kiryu$^{\textcolor[rgb]{0,0.7,0}{\spadesuit}}$, Shota Komatsu$^{\textcolor[rgb]{0,0,0.8}{\vardiamond}}$, Takuya Nishimura$^{\clubsuit}$}
\\ \vspace{1.2cm}
\footnotesize{\textit{
$^{\textcolor[rgb]{0.9,0,0}{\varheart}}$National Institute for Theoretical Physics, School of Physics and Mandelstam Institute for Theoretical Physics, University of the Witwatersrand, Johannesburg
Wits 2050, South Africa\\
\&\\
MTA Lendulet Holographic QFT Group, Wigner Research Centre, Budapest 114, P.O.B. 49, Hungary\\
$^{\textcolor[rgb]{0,0.7,0}{\spadesuit},\clubsuit}$Institute of Physics, University of Tokyo, Komaba, Meguro-ku, Tokyo 153-8902 Japan\\
$^{\textcolor[rgb]{0,0,0.8}{\vardiamond}}$Perimeter Institute for Theoretical Physics,
Waterloo, Ontario N2L 2Y5, Canada\\
\&\\
School of Natural Sciences, Institute for Advanced Study, Princeton, New Jersey 08540, USA\\
}  
\vspace{4mm}
}
\textrm{E-mail: \{minkyoo.kim$\bullet$wits.ac.za, kiryu$\bullet$hep1.c.u-tokyo.ac.jp, shota.komadze$\bullet$gmail.com, tnishimura$\bullet$hep1.c.u-tokyo.ac.jp\}  /. $\bullet\rightarrow$ @}

\par\vspace{1.5cm}

\textbf{Abstract}\vspace{2mm}
\end{center}
We study three-point functions of operators on the $1/2$ BPS Wilson loop in planar $\mathcal{N}=4$ super Yang-Mills theory. The operators we consider are ``defect changing operators'', which change the scalar coupled to the Wilson loop. We first perform the computation at two loops in general set-ups, and then study a special scaling limit called the ladders limit, in which the spectrum is known to be described by a quantum mechanics with the SL(2,$\mathbb{R}$) symmetry. In this limit, we resum the Feynman diagrams using the Schwinger-Dyson equation and determine the structure constants at all order in the rescaled coupling constant. Besides providing an interesting solvable example of defect conformal field theories, our result gives invaluable data for the integrability-based approach to the structure constants.
\noindent

\setcounter{page}{1}
\renewcommand{\thefootnote}{\arabic{footnote}}
\setcounter{footnote}{0}
\setcounter{tocdepth}{2}
\newpage
\tableofcontents

\parskip 5pt plus 1pt   \jot = 1.5ex

\newpage
\section{Introduction\label{sec:intro}}
Nonlocal operators and defects enrich our knowledge of interacting quantum field theories: In theories with mass gap, they can be important order parameters which help to distinguish different phases. The prototypical examples of such operators are the Wilson and the `t Hooft loops in gauge theories. In conformal field theories on the other hand, conformal defects lead to a new class of crossing equations, which constrain both operators in the bulk and operators on the defect\cite{BGLM,Gadde}. 

The main focus of this paper is a theory which is a gauge theory and also a conformal field theory; namely $\mathcal{N}=4$ super Yang-Mills theory ($\mathcal{N}=4$ SYM). We in particular study the three-point function of {\it defect changing operators} on the $1/2$ BPS Wilson loop.

The $1/2$ BPS Wilson loop is a supersymmetric extension of the ordinary Wilson loop, which is coupled to a scalar as well as to a gauge field\fn{${\rm Pexp}$ denotes a path-ordered exponential.},
\beq
W\equiv {\rm Tr}\left[{\rm Pexp} \left(  \oint d\tau iA_{\mu} \dot{x}^{\mu}+ \phi_i n^{i}|\dot{x}^{\mu}|\right)\right]\period
\eeq
Here $n^{i}$ is a six-dimensional unit vector, to be called the {\it R-symmetry polarization}, which  designates the scalar coupled to the Wilson loop. Local operators on the loop (denoted by $\mathcal{O}$) can be introduced by inserting fields inside the trace e.g.~
\beq
\begin{aligned}
W[\mathcal{O}]\equiv{\rm Tr}\left[{\rm Pexp}\left(  \int_{-\infty}^{\tau}\!\!\! d\tau^{\prime} iA_{\mu} \dot{x}^{\mu}+ \phi_i \textcolor[rgb]{0,0,1}{n^{i}}|\dot{x}^{\mu}|\right) \underbrace{\,\, Z^L\,\,}_{\mathcal{O}}{\rm Pexp} \left(  \int_{\tau}^{\infty}\!\!\! d\tau^{\prime\prime} iA_{\mu} \dot{x}^{\mu}+ \phi_i \textcolor[rgb]{1,0,0}{\tilde{n}^{i}}|\dot{x}^{\mu}|\right)\right]\comma
\end{aligned}
\eeq
where $Z$ is a complex scalar field.
As indicated, the polarization $n^{i}$ can change across the operator insertion.
The simplest possible operator among them is the one which has no field insertions and merely changes the polarization. We call such operators the {\it defect changing operators} (DCO).

The spectrum of such operators in the planar limit was studied extensively using integrability \cite{Drukker, CMS,GLM}. A natural next step in this direction is to compute their three-point functions\fn{Recently several structure constants on the 1/2 BPS Wilson loop were computed in \cite{CDD} by taking the limit of generic smooth Wilson loops. Also, the four-point functions of single-letter operators were computed at strong coupling using the Witten diagrams in \cite{GTR}.}. For ordinary gauge invariant operators, there exists a nonperturbative framework to study the three-point function \cite{BKV}. It is based on the fact that, in the AdS/CFT correspondence, the three-point function is dual to a closed-string world sheet whose topology is a pair of pants. The key idea in this approach is to decompose such a pair of pants into two hexagons and determine the contribution from each hexagon using integrability. 

It is then interesting to ask if we can extend this approach to more general observables involving open strings. The three-point function on the Wilson loop, which we discuss in this paper, is precisely such an observable; it corresponds to the interaction process of three open strings in AdS. With an eye toward such a direction, we perform two perturbative computations in this paper: After summarizing the set-ups and conventions in section \ref{sec:set-up}, we first compute the two-point functions of general DCO's at two loops in section \ref{sec:two-point}. Through this computation, we reproduce the anomalous dimensions computed previously in the literature. Furthermore, we also determine the normalizations of the operators, which are prerequisite for computing the scheme-independent structure constants. We then compute the three-point function at two loops in section \ref{sec:three-point}.
After doing so, we focus on a special scaling limit called the ladders limit \cite{CHMS} in section \ref{sec:ladder}, and compute the structure constants at all orders in the rescaled coupling constant using the Schwinger-Dyson equation. These results would provide important datapoints for developing the integrability-based approach in the future. In section \ref{sec:discussion}, we conclude and comment on the prospects. A few appendices are included to elucidate technical details.

\noindent{\bf Note added:} While we were writing up this article, we became aware that A.~Cavaglia, N.~Gromov, and F.~Levkovich-Maslyuk were working on a similar topic and obtained independently the results that overlap the contents of this paper. We thank them for informing us of their upcoming paper \cite{Kolyatoappear}.

\section{Set-up and conventions\label{sec:set-up}}
\subsection{Set-up for three-point functions}
Correlation functions on the 1/2 BPS straight-line Wilson loop are constrained by the ${\rm SL}(2,R)$ symmetry preserved by the Wilson loop \cite{DK}. This in particular implies that the space-time dependences of the two- and the three-point functions are completely determined. Namely, we have
\beq\label{eq:2and3}
\begin{aligned}
\langle W\left[\mathcal{O}_1(t_1)\mathcal{O}_2 (t_2)\right]\rangle &=\frac{\delta_{12}}{|t_{12}|^{2\Delta_1}}\comma\\
\langle W\left[\mathcal{O}_1(t_1)\mathcal{O}_2 (t_2)\mathcal{O}_3(t_3)\right]\rangle&=\frac{C_{123}}{|t_{12}|^{\Delta_1+\Delta_2-\Delta_3}|t_{23}|^{\Delta_2+\Delta_3-\Delta_1}|t_{31}|^{\Delta_3+\Delta_1-\Delta_2}}\comma
 \end{aligned}
 \eeq
 where $t_i$'s are positions of the operators and $t_{ij}\equiv t_i-t_j$. $\Delta_i$ and $C_{123}$ are the conformal dimension and the structure constant respectively.

\begin{figure}[t]
\centering
\includegraphics[clip,height=1.6cm]{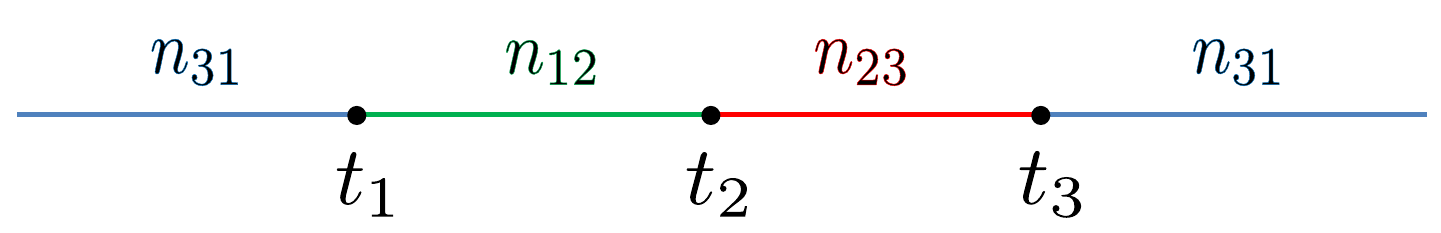} 
\caption{Three-point function of defect changing operators. Each segment is coupled to a different scalar specified by the polarizations $n_{ij}$'s.} \label{3ptconfig}
\end{figure}

As mentioned in the introduction, the main focus of this paper is the structure constant of the defect changing operators. The most general three-point functions of such operators are characterized by three six-dimensional unit vectors $n_{ij}$ parametrizing the directions of the scalars coupled to each segment of the Wilson loop (see figure \ref{3ptconfig}):
\beq
\begin{aligned}\label{setup3pt}
&\langle W\left[\mathcal{O}_1(t_1)\mathcal{O}_2 (t_2)\mathcal{O}_3(t_3)\right]\rangle\\
&\equiv \langle {\rm Tr}\left[{\rm Pexp}\left(  \int_{-\infty}^{t_1}\!\!\! d\tau \,\,iA_{\mu} \dot{x}^{\mu}+ \phi_i \textcolor[rgb]{0,0,1}{n_{31}^{i}}|\dot{x}^{\mu}|\right){\rm Pexp}\left(  \int_{t_1}^{t_2}\!\!\! d\tau \,\,iA_{\mu} \dot{x}^{\mu}+ \phi_i \textcolor[rgb]{0,0.6,0}{n_{12}^{i}}|\dot{x}^{\mu}|\right)\right.\\
&\left.\quad {\rm Pexp}\left(  \int_{t_2}^{t_3}\!\!\! d\tau\,\, iA_{\mu} \dot{x}^{\mu}+ \phi_i \textcolor[rgb]{1,0,0}{n_{23}^{i}}|\dot{x}^{\mu}|\right){\rm Pexp}\left(  \int_{t_3}^{\infty}\!\!\! d\tau \,\,iA_{\mu} \dot{x}^{\mu}+ \phi_i \textcolor[rgb]{0,0,1}{n_{31}^{i}}|\dot{x}^{\mu}|\right)\right]\rangle\period
\end{aligned}
\eeq
It is often useful to parametrize the scalar couplings by the angles between vectors $n_{ij}$ as
\beq\label{eq:defoftheta}
\cos \theta_1 \equiv n_{31}\cdot n_{12}\comma \quad \cos \theta_2 \equiv n_{12}\cdot n_{23}\comma \quad \cos \theta_3 \equiv n_{23}\cdot n_{31}\period
\eeq 
\subsection{Weak coupling expansion}
The structures of the two- and three-point functions given in \eqref{eq:2and3} apply only to properly renormalized operators. However, in the actual computation at weak coupling, we often study the correlators of un-renormalized (or equivalently bare) operators. It is therefore useful to know how to extract the conformal data from such un-renormalized correlators.

Let us first analyze the two-point functions. In general, the bare operator $\mathcal{O}^{B}$ and the renormalized operator $\mathcal{O}^{R}$ are related as\fn{Here we ignored the operator mixing since it never appears in the problems studied in this paper.}
\beq
\label{relRandB}
\mathcal{O}^{R}\equiv \frac{\epsilon^{-\gamma}}{\sqrt{a}}\mathcal{O}^{B} \comma
\eeq
where $\epsilon\sim \Lambda^{-1}$ is the cut-off, $\gamma$ is the anomalous dimension and $a$ is the finite renormalization constant needed to bring the renormalized correlator into a canonical form \eqref{eq:2and3}. Substituting \eqref{relRandB} to \eqref{eq:2and3}, we can determine the structure of the un-renormalized two-point function as
\beq\label{woo}
\langle W[\mathcal{O}^{B}(t_1)\mathcal{O}^{B}(t_2)]\rangle=\frac{a}{|t_{12}|^{2\Delta^{(0)}}}\frac{1}{\left(|t_{12}|/\epsilon\right)^{2\gamma}}\comma
\eeq
where $\Delta^{(0)}$ is the bare dimension. Both $\gamma$ and $a$ are functions of the 't Hooft coupling constant $\lambda\equiv g_{\rm YM}^2 N$, and can be expanded as
\beq
\begin{aligned}
a = 1+\lambda a^{(1)}+\lambda^2 a^{(2)} +\cdots\comma\qquad \qquad 
\gamma = \lambda \gamma ^{(1)}+ \lambda^2\gamma^{(2)}+\cdots\period
\end{aligned}
\eeq
Here we assumed that the correlator is correctly normalized at tree level, namely $\left. a\right|_{\lambda=0}=1$.
By expanding the right hand side of \eqref{woo}, we obtain the expression at weak coupling,
\beq
\begin{aligned}\label{2ptexpanded}
&\langle W[\mathcal{O}^{B}(t_1)\mathcal{O}^{B}(t_2)]\rangle=\frac{\left(1+\lambda A^{(1)}+\lambda^2A^{(2)}+\cdots\right)}{|t_{12}|^{2\Delta_{1}^{(0)}}}\comma
\end{aligned}
\eeq
with
\beq
\begin{aligned}
A^{(1)}&=a^{(1)}-2\gamma^{(1)}\log \frac{|t_{12}|}{\epsilon}\comma\\
A^{(2)}&=a^{(2)}-2a^{(1)}\gamma^{(1)}\log \frac{|t_{12}|}{\epsilon}+ 2\left(\gamma^{(1)} \log \frac{|t_{12}|}{\epsilon}\right)^2-2\gamma^{(2)}\log \frac{|t_{12}|}{\epsilon}\period
\end{aligned}
\eeq

From the relation \eqref{relRandB} and the structure of the renormalized three-point function \eqref{eq:2and3}, we can also determine the structure of the bare three-point functions at weak coupling. To simplify the expression, below we set $a^{(1)}=0$ as it is satisfied in all the examples studied in this paper. Then, using the expansion of the structure constant,
\beq
C_{123}=C_{123}^{(0)}\left(1+\lambda c_{123}^{(1)}+\lambda^2 c_{123}^{(2)}+\cdots\right)\comma
\eeq
one can write the result as
\beq\label{3ptexpanded}
\langle\mathcal{O}_1^{B}(t_1)\mathcal{O}^{B}_{2}(t_2)\mathcal{O}^{B}_{3}(t_3) \rangle=\frac{C_{123}^{(0)}\left(1+\lambda B^{(1)}+\lambda^2 B^{(2)}+\cdots\right)}{|t_{12}|^{\Delta_1^{(0)}+\Delta_2^{(0)}-\Delta_3^{(0)}}|t_{23}|^{\Delta_2^{(0)}+\Delta_3^{(0)}-\Delta_1^{(0)}}|t_{31}|^{\Delta_3^{(0)}+\Delta_1^{(0)}-\Delta_2^{(0)}}}\comma
\eeq  
with
\beq\label{Bexpanded}
\begin{aligned}
B^{(1)}=&c_{123}^{(1)}-\sum_{i}\gamma_i^{(1)}\log u_i\comma\\
B^{(2)}=&c_{123}^{(2)}+\frac{1}{2}\sum_{i}a^{(2)}_i-\sum_{i}\left(\gamma_i^{(2)}+c_{123}^{(1)}\gamma_i^{(1)}\right)\log u_i+\frac{1}{2}\sum_{i,j}\gamma_i^{(1)}\gamma_j^{(1)}\log u_i\log u_j\period
\end{aligned}
\eeq
Here $a_i$ and $\gamma_i$ are the normalization and the anomalous dimension of the operator $\mathcal{O}_i$ respectively and $u_i$ is given by
\beq
u_i \equiv \left|\frac{t_{ij}t_{ki}}{t_{jk}\epsilon}\right| 
\eeq
where $\{i,j,k\}$ is a cyclic permutation of $\{1,2,3\}$.

In the following sections, we will compare the results from perturbation with the expressions \eqref{2ptexpanded} and \eqref{3ptexpanded} and read off the anomalous dimension and the structure constant.
\subsection{Action and Propagators}
The Euclidean action\fn{Our convention is essentially the same as (42) in \cite{ESZ}. The differences are \begin{enumerate}
\item We write the action in terms of traces instead of decomposing the fields into the generators of SU($N$). 
\item We chose the Feynman gauge by setting $\xi$ in (42) of \cite{ESZ} to be $1.$
\item A sign error in front of the scalar quartic interaction in \cite{ESZ} was corrected.
\end{enumerate}} of $\mathcal{N}=4$ SYM used in this paper is
\begin{align}
S=&\frac{1}{g_{\rm YM}^2}\int d^{4} x\,\, \mathcal{L}\comma\\
\mathcal{L}=&{\rm Tr}\Big[-\frac{[D_{\mu},D_{\nu}]^2}{2} +(D_{\mu}\phi_i)^2+\frac{[\phi_i,\phi_j]^2}{2}+i\bar{\psi}\Gamma^{\mu}D_{\mu}\psi +\bar{\psi}\Gamma^{i}[\phi_i,\psi] +\del^{\mu}\bar{c}D_{\mu}c+(\del_\mu A^{\mu})^2\Big]\comma\nn
\end{align} 
with $D_{\mu}\equiv \del_{\mu} -i [A_{\mu},\quad]$. Here $c$ and $\bar{c}$ are the ghosts and $\Gamma^{A}=(\Gamma^{\mu},\Gamma^{i})$ are the ten-dimensional Dirac matrices satisfying
\beq
\tr (\Gamma^{A}\Gamma^{B})=16\delta^{AB}\period
\eeq
To emphasize that the trace here is not over the SU($N$) colour degrees of freedom, here we used the lowercase letters, ${\rm tr}$.

Using this action, one can compute the propagators as follows:
\beq
\begin{aligned}
{\rm Gluon}:&\quad\raisebox{-0.4cm}{\includegraphics[clip,height=1cm]{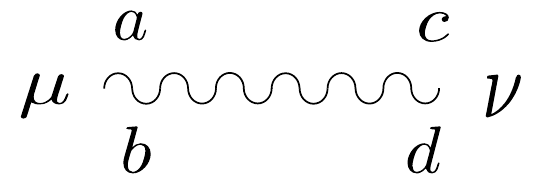}} &&=\frac{g_{\rm YM}^2 \delta^{ac}\delta^{bd}}{8\pi^2}\frac{\delta_{\mu\nu}}{|x-y|^2}\comma\\
{\rm Scalar}:&\quad\raisebox{-0.3cm}{\includegraphics[clip,height=0.9cm]{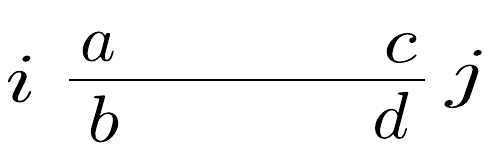}} &&=\frac{g_{\rm YM}^2 \delta^{ac}\delta^{bd}}{8\pi^2}\frac{\delta_{ij}}{|x-y|^2}\comma\\
{\rm Fermion}:&\quad\raisebox{-0.3cm}{\includegraphics[clip,height=0.9cm]{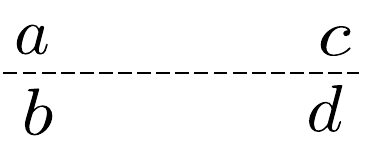}} &&=\frac{g_{\rm YM}^2 \delta^{ac}\delta^{bd}}{8\pi^2}\frac{1}{|x-y|^2}\comma\\
{\rm Ghost}:&\quad\raisebox{-0.3cm}{\includegraphics[clip,height=0.9cm]{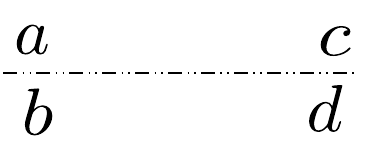}} &&=\frac{g_{\rm YM}^2 \delta^{ac}\delta^{bd}}{8\pi^2}\frac{1}{|x-y|^2}\comma
\end{aligned}
\eeq
Here $a$-$d$ are the color indices and all the propagators are proportional to $\delta^{ac}\delta^{bd}$.
To compute the loop corrections, one just need to bring down the interaction terms by expanding $e^{-S}$ and Wick-contract the fields using the propagators.

\section{Two-point functions at two loops\label{sec:two-point}}
Let us first compute the two-point function of defect changing operators at two loops. The purpose of this section is twofold; to reproduce the anomalous dimension known in the literature and to determine the normalization of the operator, which is necessary for extracting the scheme-independent structure constant from perturbative three-point functions.
\subsection{One loop}
\begin{figure}[h]
\centering
\includegraphics[clip,height=1.2cm]{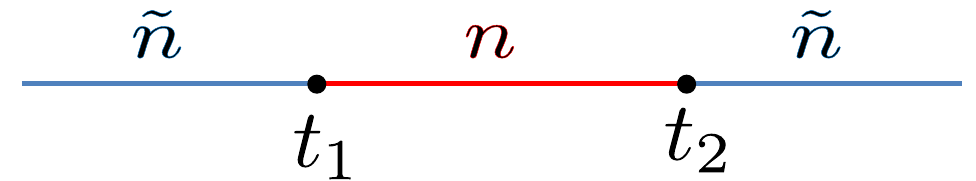} 
\caption{Two-point function of defect changing operators.} \label{2ptconfig}
\end{figure}
The two-point function we compute is of the following form (see also figure \ref{2ptconfig}):
\beq\label{setup2pt1loop}
\begin{aligned}
&\langle W\left[\mathcal{O}_1(t_1)\mathcal{O}_2 (t_2)\right]\rangle\equiv \langle {\rm Tr}\left[{\rm Pexp}\left(  \int_{-\infty}^{t_1}\!\!\! d\tau \,\,iA_{\mu} \dot{x}^{\mu}+ \phi_i \textcolor[rgb]{0,0,1}{\tilde{n}^{i}}|\dot{x}^{\mu}|\right)\right.\\
&\left.\qquad {\rm Pexp}\left(  \int_{t_1}^{t_2}\!\!\! d\tau \,\,iA_{\mu} \dot{x}^{\mu}+ \phi_i \textcolor[rgb]{1,0,0}{n^{i}}|\dot{x}^{\mu}|\right){\rm Pexp}\left(  \int_{t_2}^{\infty}\!\!\! d\tau\,\, iA_{\mu} \dot{x}^{\mu}+ \phi_i \textcolor[rgb]{0,0,1}{\tilde{n}^{i}}|\dot{x}^{\mu}|\right)\right]\rangle\period
\end{aligned}
\eeq
Owing to the SO(6) invariance, it only depends on the inner product $n\cdot \tilde{n}$ and the coupling constant $\lambda$. To perform the perturbative computation, we just need to expand the exponentials in \eqref{setup2pt1loop} and contract them with propagators and vertices.

\begin{figure}[t]
\centering
\includegraphics[clip,height=1.2cm]{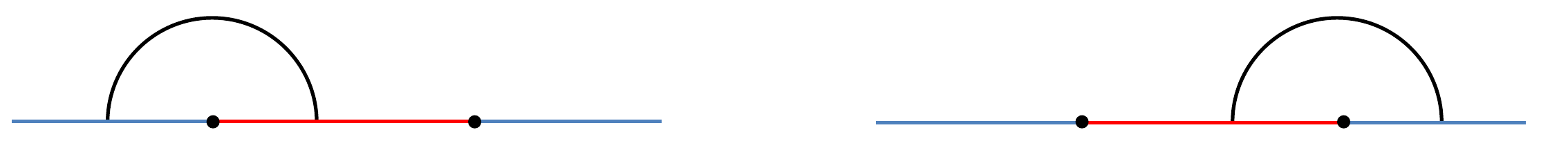} 
\caption{One-loop corrections to the two-point function of defect changing operators. Owing to the cancellation between gluons and scalars, the only diagrams that survive are the ones in which the propagator connects two different segments.} \label{2pt1loop}
\end{figure}

At one loop, one can only have a single propagator (without vertices). The propagator can be either a gauge field or a scalar field and takes the form,
\beq\label{gaugescalarpr}
\text{Gauge}: -\frac{\lambda }{8\pi^2}\frac{1}{\tau_{12}^2}\comma\qquad \text{Scalar}: (n_1\cdot n_2)\frac{\lambda}{8\pi^2} \frac{1}{\tau_{12}^2} \comma
\eeq
where $\tau_{12}=\tau_1-\tau_2$ is the distance between two end points and $n_{1}$ and $n_2$ (which can be either $n$ or $\tilde{n}$) are the polarization vectors at each end point. The extra minus sign for gluons comes from factors of $i$ in the exponentials in \eqref{setup2pt1loop}. As is clear from \eqref{gaugescalarpr}, the contributions from the gauge field and the scalar field cancel out if $n_1=n_2$, since both $n$ and $\tilde{n}$ have a unit norm. We are thus left with diagrams in which a propagator connects two segments with different polarizations (see figure \ref{2pt1loop}).

To compute such diagrams, we introduce a UV regularization by cutting out a small circle of radius\footnote{The factor of $1/2$ is just a useful convention which simplifies the final result.} $\epsilon/2$ around each operator. This leads to a regularized integral
\beq
\begin{aligned}
\langle W\!\left[\mathcal{O}_1(t_1)\mathcal{O}_2 (t_2)\right]\rangle_{\text{1-loop}} =\frac{\lambda(n\cdot \tilde{n}-1)}{8\pi^{2}}\left( \int^{t_{1}^{-}}_{-\infty}d\tau_{1}\int^{t_{2}^{-}}_{t_{1}^{+}}d\tau_{2}\frac{1}{\tau_{12}^{2}} +\int^{t_{2}^{-}}_{t_{1}^{+}}d\tau_{1}\int^{\infty}_{t_{2}^{+}}d\tau_{2}\frac{1}{\tau_{12}^{2}} \right)\comma
\end{aligned}
\eeq
with $t_{i}^{\pm}=t_i \pm \epsilon/2$. This integral can be easily evaluated using
\beq
\int_{d}^{c}d\tau_1\int^{a}_{b} d\tau_2 \,\frac{1}{\tau_{12}^2}=\log \frac{(a-c)(b-d)}{(a-d)(b-c)}\comma
\eeq
as
\beq
\langle W\!\left[\mathcal{O}_1(t_1)\mathcal{O}_2 (t_2)\right]\rangle_{\text{1-loop}} =\frac{\lambda(n\cdot \tilde{n}-1)}{4\pi^{2}}\log \frac{t_{12}}{\epsilon}+O(\epsilon)\period
\eeq
Comparing this with \eqref{2ptexpanded}, we can determine the one-loop normalization $a^{(1)}$ and the anomalous dimension $\gamma^{(1)}$ as
\begin{align}
a^{(1)}=0\comma\qquad \gamma^{(1)}=\frac{1-n\cdot \tilde{n}}{8\pi^{2}}\period\label{1-loop2pt}
\end{align}
The result for $\gamma^{(1)}$ of course matches the one in the literature \cite{DF}, but it also shows that the normalization at one loop $a^{(1)}$ vanishes in our scheme.

\subsection{Two loops\label{subsec:2pt2loop}}
Let us now proceed to the two-loop computation. At two loops, there appear three types of diagrams; the ladder, the vertex and the self-energy. Below we are going to evaluate them one by one. 
\subsubsection*{Ladder diagrams}
The first diagrams are the ladder diagrams, which consist only of propagators. For this class of diagrams, the computation proceeds in a similar manner as in the previous subsection. Also here, the diagrams that contain propagators connecting the same segment vanish due to the cancellation between the scalar and the gluon.
\begin{figure}[t]
\centering
\includegraphics[clip,height=2cm]{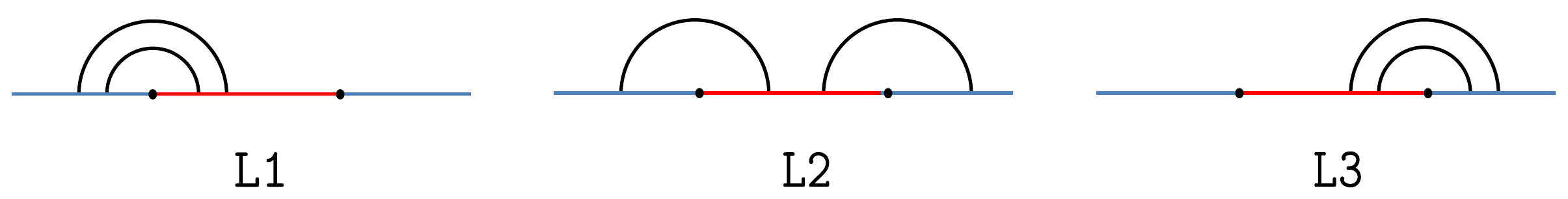} 
\caption{Ladder diagrams that contribute to the two-point function at two loops. Here thick black curves represent either a scalar propagator or a gluon propagator.} \label{2ptladder2loop}
\end{figure}

Thus only non-zero diagrams are the ones depicted in figure \ref{2ptladder2loop}, which are given by
\beq
\begin{aligned}
{\tt L1}=&\left(\frac{(n\cdot \tilde{n}-1)\lambda}{8\pi^{2}}\right)^{2}\int^{t_1^{-}}_{-\infty} d\tau_1\int^{t_1^{-}}_{\tau_1} d\tau_2\int^{t_2^{-}}_{t_{1}^{+}}d\tau_3\int^{t_2^{-}}_{\tau_3}d\tau_{4}\,\,\frac{1}{\tau_{14}^2}\frac{1}{\tau_{23}^2}
\comma\\
{\tt L2}=&\left(\frac{(n\cdot \tilde{n}-1)\lambda}{8\pi^{2}}\right)^{2}\int^{t_1^{-}}_{-\infty} d\tau_1\int^{t_{2}^{-}}_{t_1^{+}} d\tau_2\int^{t_{2}^{-}}_{\tau_{2}}d\tau_3\int^{\infty}_{t_2^{+}}d\tau_{4}\,\,\frac{1}{\tau_{12}^2}\frac{1}{\tau_{34}^2}
\comma
\\
{\tt L3}=&\left(\frac{(n\cdot \tilde{n}-1)\lambda}{8\pi^{2}}\right)^{2}\int^{t_2^{-}}_{t_{1}^{+}} d\tau_1\int^{t_2^{-}}_{\tau_1} d\tau_2\int_{t_2^{+}}^{\infty}d\tau_3\int^{\infty}_{\tau_3}d\tau_{4}\,\,\frac{1}{\tau_{14}^2}\frac{1}{\tau_{23}^2}
\period
\end{aligned}
\eeq
Each of these integrals can be evaluated straightforwardly as follows\footnote{The equivalence between ${\tt L1}$ and ${\tt L3}$ can be shown at the level of integrands by performing the translation $\tau_i \to \tau_i -c$ and the reflection $\tau_i \to -\tau_i$.}:
\beq
\begin{aligned}
{\tt L1}={\tt L3}&=\left(\frac{(n\cdot \tilde{n}-1)\lambda}{8\pi^{2}}\right)^{2}\left[\frac{\pi^2}{6}-\log\frac{t_{21}}{\epsilon}+\frac{1}{2}\left(\log\frac{t_{21}}{\epsilon}\right)^2\right]+O(\epsilon)\comma\\
{\tt L2}&=\left(\frac{(n\cdot \tilde{n}-1)\lambda}{8\pi^{2}}\right)^{2}\left[-\frac{\pi^2}{6}+\left(\log\frac{t_{21}}{\epsilon}\right)^2\right]+O(\epsilon)\period
\end{aligned}
\eeq
Summing three terms, we get
\begin{align}
{\tt L}\equiv {\tt L1}+{\tt L2}+{\tt L3}=\left(\frac{(n\cdot \tilde{n}-1)\lambda}{8\pi^{2}}\right)^{2}\left[\frac{\pi^2}{6}-2\log\frac{t_{21}}{\epsilon}+2\left(\log\frac{t_{21}}{\epsilon}\right)^2\right]\period\label{twoladder}
\end{align}
\subsubsection*{Vertex diagrams}
The second diagrams are the ones which contain one interaction vertex. Written explicitly, they arise from the Wick contraction of the following terms: 
\beq
\begin{aligned}
\frac{i^{3}}{3!}\int d\tau_{1}d\tau_{2}d\tau_{3}\left<{\rm Tr}\,{\rm P}[\mathcal{A}(\tau_{1})\mathcal{A}(\tau_{2})\mathcal{A}(\tau_{3})]\left(\frac{2i}{g_{\rm YM}^{2}}\int d^{4}x {\rm Tr}\{ \partial_{\mu}A_{\nu}(x)[A_{\mu}(x),A_{\nu}(x)]\}\right)\right>\\
+\frac{i}{2!1!}\int d\tau_{1}d\tau_{2}d\tau_{3}\left<{\rm Tr}\,{\rm P}[\Phi_1(\tau_{1})\Phi_2(\tau_{2})\mathcal{A}(\tau_{3})]\left(\frac{2i}{g_{\rm YM}^{2}}\int d^{4}x{\rm Tr}\,\{\partial_{\mu}\phi(x)\left[A_{\mu}(x),\phi(x)\right]\}\right)\right>\period\nn
\end{aligned}
\eeq
Here $\mathcal{A}\equiv A_{\mu}\dot{x}^{\mu}$ and $\Phi_i\equiv (\phi\cdot n_i) |\dot{x}|$ with $n_{1,2}$ being the polarization vectors at each end-point. 
\begin{figure}[t]
\centering
\includegraphics[clip,height=6.3cm]{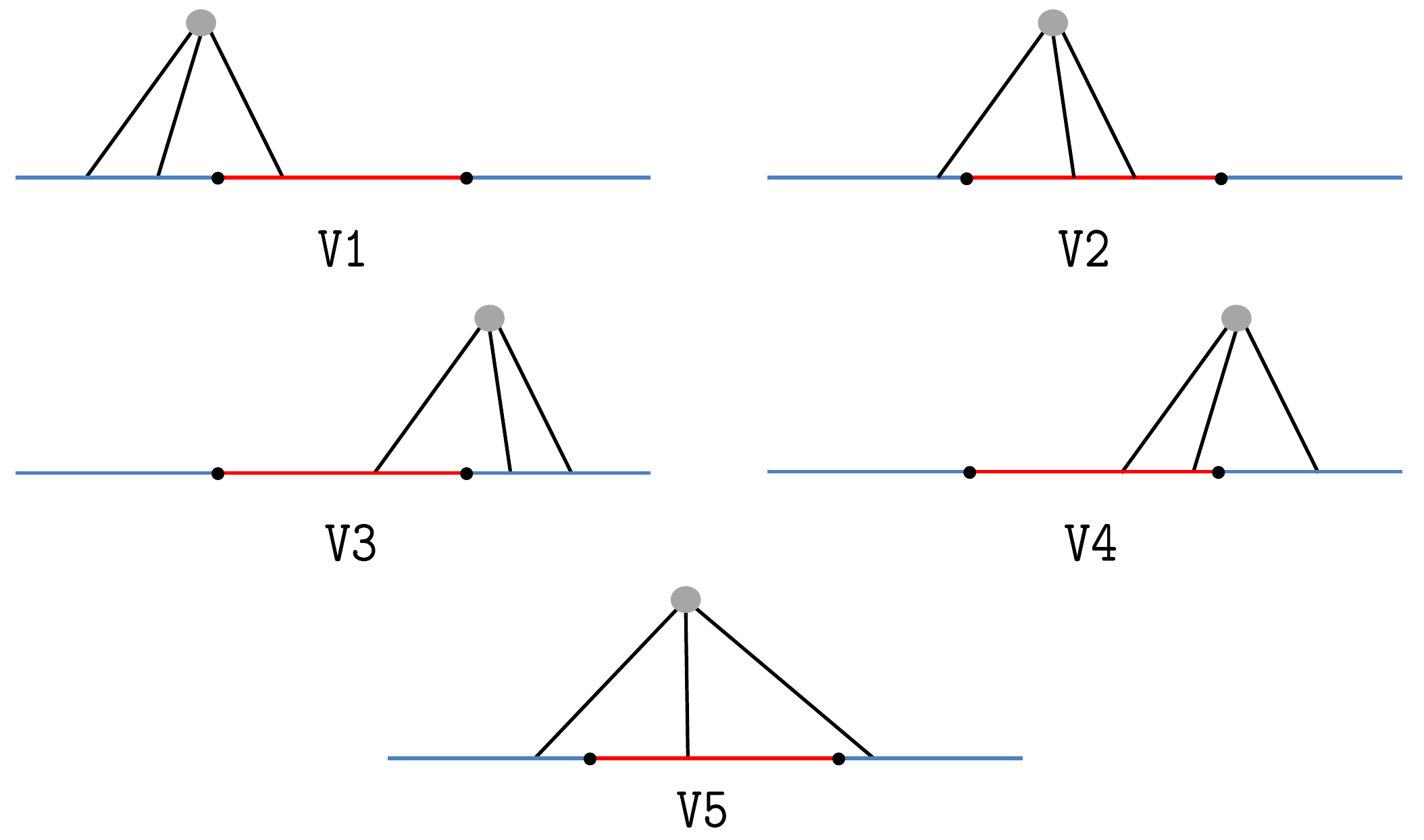} 
\caption{The two-loop diagrams for the two-point function that involve one interaction vertex. Here again, each black line can be either scalar or gluon.} \label{2ptvertex}
\end{figure}

The Wick contraction of the above correlator leads to a set of diagrams shown in figure \ref{2ptvertex}. To illustrate how the computation goes, let us focus on the diagram ${\tt V1}$. In the diagram ${\tt V1}$, the contribution from the scalar-scalar-gauge vertex consists of three different terms depending on the path-ordering:
\beq\label{V1onetwothree}
\begin{aligned}
i \int_{\substack{\tau_1,\tau_2\in [-\infty,t_1]\\\tau_1<\tau_2}} d\tau_1 d\tau_{2}\int_{\tau_3\in [t_1,t_2]}d\tau_{3}\left[\left<{\rm Tr}\left(\mathcal{A}(\tau_3)\tilde{\Phi}(\tau_2)\tilde{\Phi}(\tau_1)\right)\right>\right.\\
\left.+\left<{\rm Tr}\left(\Phi(\tau_3)\mathcal{A}(\tau_2)\tilde{\Phi}(\tau_1)\right)\right>+\left<{\rm Tr}\left(\Phi(\tau_3)\tilde{\Phi}(\tau_2)\mathcal{A}(\tau_1)\right)\right>\right]\period
\end{aligned}
\eeq
Here $\Phi\equiv n\cdot \phi$ and $\tilde{\Phi}\equiv \tilde{n}\cdot \phi$, and we did not write the interaction vertex for brevity. Among these three terms, the first term, which has two scalars in the segment $[-\infty,t_1]$, does not contribute to the final answer since it is proportional to $\tilde{n}\cdot \tilde{n}=1$ and is cancelled precisely by a similar contribution from the three-gauge vertex. On the other hand, from the second term we get
\beq\label{secondterminV1}
(\text{second term})= -\frac{\lambda^2 (n\cdot \tilde{n})}{4(4\pi^2)^3}\int_{\substack{\tau_1,\tau_2\in [-\infty,t_1]\\\tau_1<\tau_2}} d\tau_1d\tau_2 \int_{\tau_3\in [t_1,t_2]}d\tau_3 \left(-\del_{\tau_1}Y_{123}+\del_{\tau_3}Y_{123}\right)\comma
\eeq
with (see Appendix \ref{Y123}  for more details)
\beq\label{Y123explicit}
Y_{123}\left(\equiv \int \frac{d^{4}x_5}{x_{15}^2x_{25}^2x_{35}^2}\right)=-2\pi ^2 \left(\frac{\log |\tau_{12}|}{\tau_{13}\tau_{23}}+\frac{\log |\tau_{13}|}{\tau_{12}\tau_{32}}+\frac{\log |\tau_{23}|}{\tau_{21}\tau_{31}}\right)\period
\eeq
In \eqref{secondterminV1}, the term $-\del_{\tau_1}Y_{123}$ comes from the contraction with the interaction $\int d^4 x \,\text{Tr} (\del_{\mu}\phi  A_{\mu} \phi)$ while the term $\del_{\tau_3}Y_{123}$ comes from the contraction with $-\int d^4 x \,\text{Tr} (\del_{\mu}\phi \phi A_{\mu} )$. Similarly the third term \eqref{V1onetwothree} yields\fn{Here $-\del_{\tau_3}Y_{123}$ comes from the contraction with $\int d^4 x \,\text{Tr} (\del_{\mu}\phi  A_{\mu} \phi)$ while $\del_{\tau_1}Y_{123}$ comes from the contraction with $-\int d^4 x \,\text{Tr} (\del_{\mu}\phi \phi A_{\mu} )$.}
\beq\label{thirdterminV1}
(\text{third term})= -\frac{\lambda^2 (n\cdot \tilde{n})}{4(4\pi^2)^3}\int_{\substack{\tau_1,\tau_2\in [-\infty,t_1]\\\tau_1<\tau_2}} d\tau_1d\tau_2 \int_{\tau_3\in [t_1,t_2]}d\tau_3 \left(-\del_{\tau_3}Y_{123}+\del_{\tau_2}Y_{123}\right)\period
\eeq
Adding up the two terms, \eqref{secondterminV1} and \eqref{thirdterminV1}, and also the contributions from the three-gauge vertex, we arrive at the following result for the diagram ${\tt V1}$: 
\beq
\begin{aligned}
{\tt V1}= -\frac{\lambda^2 (n\cdot \tilde{n}-1)}{4(4\pi^2)^3}\int_{-\infty}^{t_1^{-}}d\tau_1\int_{-\infty}^{t_1^{-}}d\tau_2\int_{t_1^{+}}^{t_2^{-}}d\tau_3 \,\epsilon (\tau_1-\tau_2) \del_{\tau_1} Y_{123}\period
\end{aligned}
\eeq
Here we used the permutation symmetry of $Y_{123}$, $Y_{123}=Y_{213}$ etc., to simplify the result and $\epsilon (x)\equiv \theta (x)-\theta (-x)$ with $\theta (x)$ being the step function.

By performing the similar analysis, we arrive at the following results for other diagrams:
\beq
\begin{aligned}
{\tt V2}=&-\frac{\lambda^2 (n\cdot \tilde{n}-1)}{4(4\pi^2)^3}\int_{-\infty}^{t_{1}^{-}}d\tau_{1}\int_{t_{1}^{+}}^{t_{2}^{-}}d\tau_{2}\int_{t_{1}^{+}}^{t_{2}^{-}}d\tau_{3}\,\epsilon(\tau_{2}-\tau_{3})\del_{\tau_2}Y_{123}\comma\\
{\tt V3}=&-\frac{\lambda^2 (n\cdot \tilde{n}-1)}{4(4\pi^2)^3}\int_{t_{1}^{+}}^{t_{2}^{-}}d\tau_{1}\int_{t_{2}^{+}}^{\infty}d\tau_{2}\int_{t_{2}^{+}}^{\infty}d\tau_{3}\,\epsilon(\tau_{2}-\tau_{3})\del_{\tau_2}Y_{123}\comma\\
{\tt V4}=&-\frac{\lambda^2 (n\cdot \tilde{n}-1)}{4(4\pi^2)^3}\int_{t_{1}^{+}}^{t_{2}^{-}}d\tau_{1}\int_{t_{1}^{+}}^{t_{2}^{-}}d\tau_{2}\int_{t_{2}^{+}}^{\infty}d\tau_{3}\,\epsilon(\tau_{1}-\tau_{2})\del_{\tau_1}Y_{123}\comma\\
{\tt V5}=&-\frac{\lambda^2 (n\cdot \tilde{n}-1)}{4(4\pi^2)^3}\int_{-\infty}^{t_{1}^{-}}d\tau_{1}\int_{t_{1}^{+}}^{t_{2}^{-}}d\tau_{2}\int_{t_{2}^{+}}^{\infty}d\tau_{3}\left(\del_{\tau_1}Y_{123}-\del_{\tau_3}Y_{123}\right)\period
\end{aligned}
\eeq
To proceed, we perform the integration by parts to each contribution and rewrite them using $\del_x \epsilon (x)=2\delta(x)$ as
\begin{align}
{\tt V1}=& -\frac{\lambda^2 (n\cdot \tilde{n}-1)}{4(4\pi^2)^3}\left[\int_{-\infty}^{t_1^{-}}d\tau_2\int_{t_1^{+}}^{t_2^{-}}d\tau_3 \,  Y_{t_1^{-}23}
-2\int_{-\infty}^{t_1^{-}}d\tau_1\int_{-\infty}^{t_1^{-}}d\tau_2\int_{t_1^{+}}^{t_2^{-}}d\tau_3 \,\delta (\tau_1-\tau_2)  Y_{123}\right]\nn\\
=&-\frac{\lambda^2 (n\cdot \tilde{n}-1)}{4(4\pi^2)^3}\left[\int_{-\infty}^{t_1^{-}}d\tau_2\int_{t_1^{+}}^{t_2^{-}}d\tau_3 \,  Y_{t_1^{-}23}
-2\int_{-\infty}^{t_1^{-}}d\tau_2\int_{t_1^{+}}^{t_2^{-}}d\tau_3 \,  Y_{223}\right]\period
\end{align}
We thus get
\beq\label{vertexsum}
\begin{aligned}
{\tt V}\equiv&{\tt V1}+{\tt V2}+{\tt V3}+{\tt V4}+{\tt V5}+{\tt V6}=\\
&-\frac{\lambda^2 (n\cdot \tilde{n}-1)}{4(4\pi^2)^3}\left[\int_{-\infty}^{t_1^{-}}d\tau_2\int_{t_1^{+}}^{t_2^{-}}d\tau_3 \,  \left(Y_{t_1^{-}23}+Y_{t_1^{+}23}+Y_{t_2^{-}23}+Y_{t_2^{+}23}\right)\right.\\
&+\int_{t_{1}^{+}}^{t_{2}^{-}}d\tau_{2}\int_{t_{2}^{+}}^{\infty}d\tau_{3}\left(Y_{t_1^{-}23}+Y_{t_1^{+}23}+Y_{t_2^{-}23}+Y_{t_2^{+}23}\right)\\
&\left.-2\int_{-\infty}^{t_1^{-}}d\tau_2\int_{t_1^{+}}^{t_2^{-}}d\tau_3 \,  (Y_{223}+Y_{233})-2\int_{t_{1}^{+}}^{t_{2}^{-}}d\tau_{2}\int_{t_{2}^{+}}^{\infty}d\tau_{3}\,  (Y_{223}+Y_{233})\right]\period
\end{aligned}
\eeq
Note that the last line in this expression contains the function $Y$ evaluated at the coincident points and is therefore divergent. A convenient way to regularize these integrals is to use the dimensional regularization\fn{For derivation of \eqref{Yregularized}, see for instance Appendix A of \cite{Giombiloop}.}, which renders $Y$ to be
\beq\label{Yregularized}
\begin{aligned}
Y^{\epsilon}_{223}&=\int  \frac{d^{4-\epsilon}x_5}{x_{25}^4x_{35}^2}=\frac{\pi^{2-\frac{\epsilon}{2}}}{x_{23}^{2-\epsilon}}\frac{\Gamma \left(1-\frac{\epsilon}{2}\right)\Gamma \left(-\frac{\epsilon}{2}\right)\Gamma \left(1+\frac{\epsilon}{2}\right)}{\Gamma (1-\epsilon)}\comma\\
Y^{\epsilon}_{233}&=\int  \frac{d^{4-\epsilon}x_5}{x_{25}^2x_{35}^4}=Y^{\epsilon}_{223}\period
\end{aligned}
\eeq
\subsubsection*{Self-energy diagrams}
\begin{figure}[t]
\centering
\includegraphics[clip,height=1.3cm]{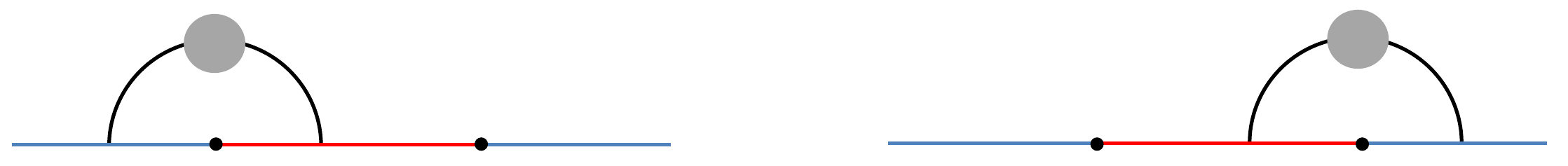} 
\caption{The self-energy diagrams that contribute to the two-loop two-point function. The sum of these two diagrams is given by \eqref{Sself}.} \label{self}
\end{figure}
We now discuss the contribution from the self-energy diagrams (see figure \ref{self}). The one-loop correction to the gauge and the scalar propagators were already computed in the literature and the result in the dimensional regularization reads \cite{DP}
\beq
\begin{aligned}
\raisebox{-0.3cm}{\includegraphics[clip,height=0.9cm]{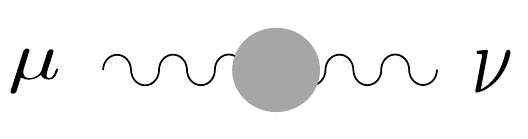}} &=\frac{g_{\rm YM}^4\delta^{ac}\delta^{bd} \delta_{\mu\nu}}{2(4\pi^2)^3}(Y^{\epsilon}_{223}+Y^{\epsilon}_{233})\comma\\
\raisebox{-0.3cm}{\includegraphics[clip,height=0.9cm]{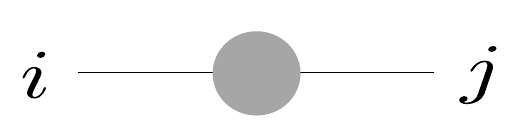}} &=-\frac{g_{\rm YM}^4 \delta^{ac}\delta^{bd}\delta_{ij}}{2(4\pi^2)^3}(Y^{\epsilon}_{223}+Y^{\epsilon}_{233})\period
\end{aligned}
\eeq
Here again $\delta^{ac}\delta^{bd}$ is the color factor\fn{Here we did not write them on the left hand side for simplicity.}.
Using these corrected propagators, one can compute the contribution from the self-energy diagrams as follows:
\beq\label{Sself}
{\tt S}=-\frac{\lambda^2 (n\cdot \tilde{n}-1)}{2(4\pi^2)^3}\left[\int_{-\infty}^{t_1^{-}}d\tau_2\int_{t_1^{+}}^{t_2^{-}}d\tau_3 \,  (Y^{\epsilon}_{223}+Y^{\epsilon}_{233})+\int_{t_{1}^{+}}^{t_{2}^{-}}d\tau_{2}\int_{t_{2}^{+}}^{\infty}d\tau_{3}\,  (Y^{\epsilon}_{223}+Y^{\epsilon}_{233})\right]\period
\eeq
It is then easy to verify that the contribution from the self-energy diagrams precisely cancels the divergent terms in \eqref{vertexsum}. 

Using the expression for $Y_{123}$ \eqref{Y123explicit}, one can straightforwardly evaluate the remaining integral\fn{In terms of polylogarithms.} to get
\beq\label{VandS}
{\tt V}+{\tt S}=-2(n\cdot \tilde{n}-1)\left(\frac{\lambda}{8\pi^{2}}\right)^2\left(\frac{\pi^{2}}{3}\log\frac{t_{21}}{\epsilon}+3\zeta(3)\right)\period
\eeq
\subsubsection*{Final result}
Now by summing up all the contributions \eqref{twoladder} and \eqref{VandS}, we get the result for the two-point function at two loops:
\beq
\begin{aligned}
\langle W\!\left[\mathcal{O}_1(t_1)\mathcal{O}_2 (t_2)\right]\rangle_{\text{2-loop}} =\left(\frac{\lambda}{8\pi^2}\right)^2\left[(n\cdot \tilde{n}-1)^2\left(\frac{\pi^2}{6}-2\log\frac{t_{21}}{\epsilon}+2\left(\log\frac{t_{21}}{\epsilon}\right)^2\right)\right.\\
\left.-2(n\cdot \tilde{n}-1)\left(\frac{\pi^{2}}{3}\log\frac{t_{21}}{\epsilon}+3\zeta(3)\right)\right]\period
\end{aligned}
\eeq
By comparing the weak coupling expansion of the two-point function \eqref{Bexpanded}, one can finally obtain the two-loop anomalous dimension $\gamma^{(2)}_{j}$ and the constant term $a^{(2)}_{j}$ as follows:
\begin{align}
a^{(2)}&=\frac{1}{(8\pi^{2})^{2}}\Big[\frac{\pi^{2}}{12}(n\cdot \tilde{n}-1)^{2}-3\zeta(3)(n\cdot \tilde{n}-1)\Big],\label{2ptconst}\\
\gamma^{(2)}&=\frac{1}{(8\pi^{2})^{2}}\Big[(n\cdot \tilde{n}-1)^{2}+\frac{\pi^{2}}{3}(n\cdot \tilde{n}-1)\Big].\label{2ptanomalous}
\end{align}
Again, the result for $\gamma^{(2)}$ matches the one in the literature \cite{DF}.

\section{Three-point functions at two loops\label{sec:three-point}}
We now set out to compute the three-point functions of DCO's on the Wilson line, given explicitly in \eqref{setup3pt}. The strategy of the computation is essentially the same as in the previous section; we list up all possible diagrams and compute each integral explicitly by using appropriate regularisations. Of course, this is easier said than done; the number of diagrams that contribute at a given loop order proliferate as we increase the number of operators.

To circumvent this complication, we use the following trick: When the polarizations of two segments are identical, the quantum correction involving these two segments must vanish owing to the supersymmetry. This implies that the polarization vectors $n_{ij}$'s enter in the final result only through the combinations\fn{At one loop, the result is linear in such combinations while it consists of linear and quadratic pieces at two loops.
} $(n_{ij}\cdot n_{kl}-1)$. In addition, the final result must be symmetric with respect to the permutation of the operator labels, $1,2$ and $3$. Therefore, instead of computing all possible diagrams, one can just focus on the coefficients of certain monomials of $(n_{ij}\cdot n_{kl}-1)$'s, and symmetrize the resulting expression with respect to the permutation of the operators to get the full answer. 
\subsection{One loop}
\begin{figure}[t]
\centering
\includegraphics[clip,height=1.3cm]{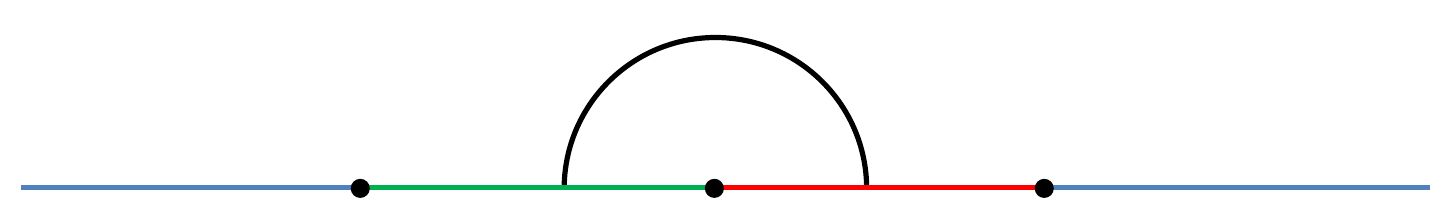} 
\caption{The one-loop diagram for the three-point function which is proportional to $(n_{12}\cdot n_{23}-1)$. The result of the computation is given in \eqref{3pt1loopevaluated}.} \label{3pt1loop}
\end{figure}
At one loop, the result is linear in $(n_{ij}\cdot n_{kl}-1)$. Therefore by using the trick explained above, we can just focus on the term proportional to $(n_{12}\cdot n_{23}-1)$ and symmetrize the result to get the full expression. As shown in figure \ref{3pt1loop}, there is only one type of diagrams that produce this contribution. As in the case of the two-point function, one can straightforwardly evaluate them as
\beq\label{3pt1loopevaluated}
\begin{aligned}
\frac{\lambda(n_{12}\cdot n_{23}-1)}{8\pi^{2}}\int^{t_2^{-}}_{t_1^{+}}d\tau_1 \int_{t_2^{+}}^{t_3^{-}}d\tau_2\,\frac{1}{\tau_{12}^2}= 
\frac{\lambda(n_{12}\cdot n_{23}-1)}{8\pi^{2}}\log\frac{t_{21}t_{32}}{t_{31}\epsilon}
\end{aligned}
\eeq

After the symmetrization, we get the full answer,
\begin{equation}
\langle W\!\left[\mathcal{O}_1(t_1)\mathcal{O}_2 (t_2)\mathcal{O}_3 (t_3)\right]\rangle_{\text{1-loop}} =\frac{\lambda}{8\pi^{2}}\sum_{\{i,j,k\}}(n_{ij}\cdot n_{jk} -1)\log\left|\frac{t_{ij}t_{jk}}{t_{ki}\epsilon}\right|,
\end{equation}
where the sum is over $\{i,j,k\}=\{1,2,3\},\{2,3,1\},\{3,1,2\}$. Comparing this result with the weak coupling expansion given in \eqref{Bexpanded}, one can read off the anomalous dimension and the structure constant as\fn{Here we already used the fact that the one-loop normalization $a^{(1)}$ vanishes in our scheme. See \eqref{1-loop2pt}.}
\begin{align*}
&\gamma_{j}^{(1)}=\frac{(1-n_{ij}\cdot n_{jk} )}{8\pi^{2}}\comma\qquad  c_{123}^{(1)}=0.
\end{align*}
As expected the result for the anomalous dimension matches the previous result \eqref{1-loop2pt}. This also shows that the one-loop structure constant is exactly zero. 
\subsection{Two loops}
At two loops, the result consists of terms quadratic in $(n_{ij}\cdot n_{jk}-1)$ and terms linear in $(n_{ij}\cdot n_{jk}-1)$. The quadratic terms come from the ladder-like diagrams while the linear terms arise from the vertex diagrams and the self-energy diagrams.

\begin{figure}[t]
\begin{minipage}{0.4\hsize}
\centering
\raisebox{-1cm}{\includegraphics[clip,height=0.8cm]{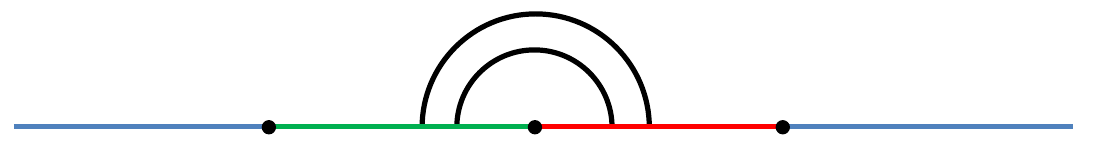}}
\end{minipage}
\begin{minipage}{0.55\hsize}
\centering
\includegraphics[clip,height=0.9cm]{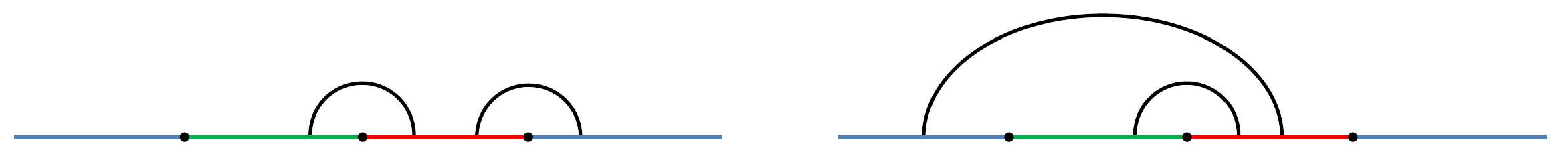}
\end{minipage}\vspace{0.1cm}\\
\begin{minipage}{0.4\hsize}
\centering
$(n_{12}\cdot n_{23}-1)^2$
\end{minipage}
\begin{minipage}{0.55\hsize}
\centering
$(n_{12}\cdot n_{23}-1)(n_{23}\cdot n_{31}-1)$
\end{minipage}
\caption{The two-loop ladder diagrams. The left diagram produces the term proportional to  $(n_{12}\cdot n_{23}-1)^2$ while the right two diagrams produce the term proportional to $(n_{12}\cdot n_{23}-1)(n_{23}\cdot n_{31}-1)$.} \label{3ptladder2loop}
\end{figure}

Let us first compute the quadratic terms. For this purpose, it is enough to compute the terms proportional to $(n_{12}\cdot n_{23}-1)^{2}$ and $(n_{12}\cdot n_{23}-1)(n_{23}\cdot n_{31}-1)$. The diagrams that contribute to these two terms are given in figure \ref{3ptladder2loop}. Then, the term proportional to $(n_{12}\cdot n_{23}-1)^{2}$ can be computed as
\beq
\begin{aligned}
&\left(\frac{(n_{12}\cdot n_{23}-1)\lambda}{8\pi^{2}}\right)^{2}\int^{t_2^{-}}_{t_1^{+}} d\tau_1\int^{t_2^{-}}_{\tau_1} d\tau_2\int^{t_3^{-}}_{t_{2}^{+}}d\tau_3\int^{t_3^{-}}_{\tau_3}d\tau_{4}\,\,\frac{1}{\tau_{14}^2}\frac{1}{\tau_{23}^2}\\
&=\left(\frac{(n_{12}\cdot n_{23}-1)\lambda}{8\pi^{2}}\right)^2\left[\frac{\pi^{2}}{6}-\log\left|\frac{t_{12}t_{23}}{t_{31}\epsilon}\right|+\frac{1}{2}\left(\log\left|\frac{t_{12}t_{23}}{t_{31}\epsilon}\right|\right)^2\right]\period
\end{aligned}
\eeq
On the other hand, the term proportional to $(n_{12}\cdot n_{23}-1)(n_{23}\cdot n_{31}-1)$ is given by
\beq
\begin{aligned}
&(n_{12}\cdot n_{23}-1)(n_{23}\cdot n_{31}-1)\left(\frac{\lambda}{8\pi^{2}}\right)^{2}\left[\int^{t_2^{-}}_{t_1^{+}} d\tau_1\int^{t_{3}^{-}}_{t_2^{+}} d\tau_2\int^{t_{3}^{-}}_{\tau_{2}}d\tau_3\int^{\infty}_{t_3^{+}}d\tau_{4}\,\,\frac{1}{\tau_{12}^2}\frac{1}{\tau_{34}^2}\right.\\
&\left.\hspace{6cm} +\int^{t_1^{-}}_{-\infty} d\tau_1\int^{t_{2}^{-}}_{t_1^{+}} d\tau_2\int^{t_{3}^{+}}_{t_2^{+}}d\tau_3\int^{t_3^{+}}_{\tau_3}d\tau_{4}\,\,\frac{1}{\tau_{14}^2}\frac{1}{\tau_{23}^2}\right]\\
&=-(n_{12}\cdot n_{23}-1)(n_{23}\cdot n_{31}-1)\left(\frac{\lambda}{8\pi^{2}}\right)^2\left(\frac{\pi^{2}}{6}-\log\left|\frac{t_{12}t_{23}}{t_{31}\epsilon}\right|\log\left|\frac{t_{23}t_{31}}{t_{12}\epsilon}\right|\right)\period
\end{aligned}
\eeq

\begin{figure}[t]
\centering
\includegraphics[clip,height=5cm]{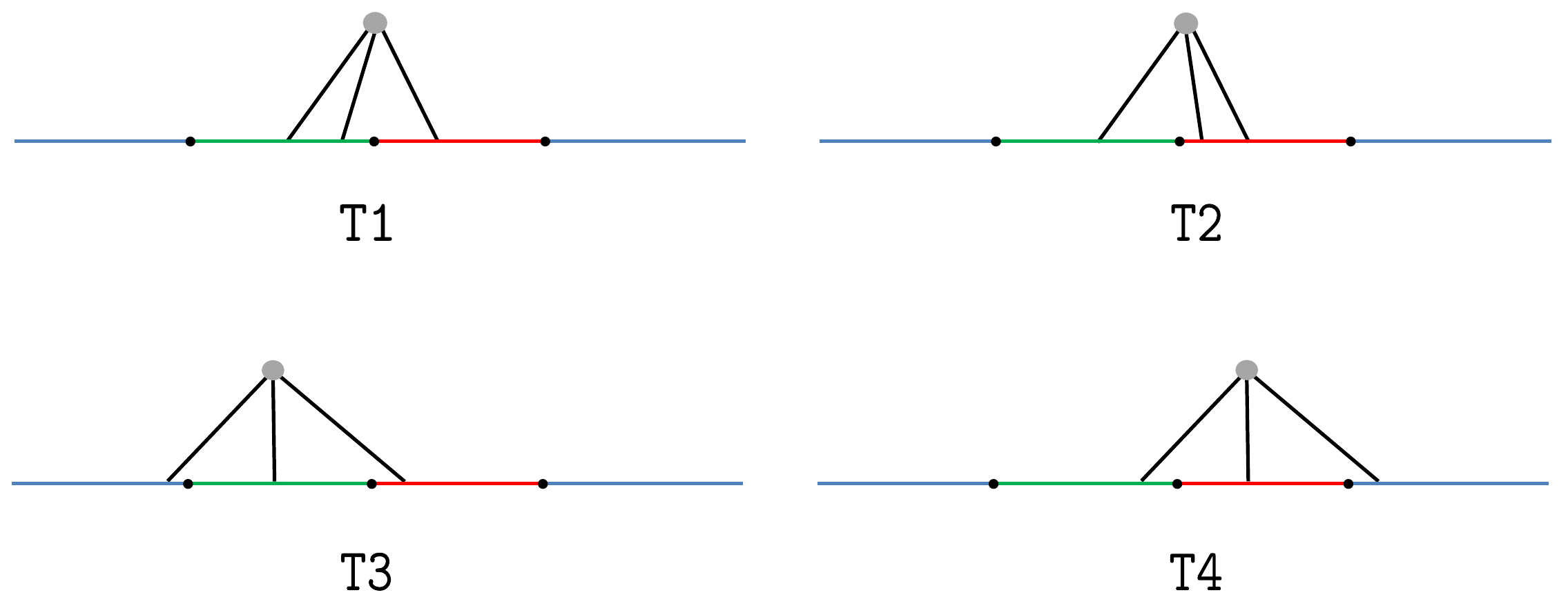}
\caption{The two-loop diagram with one interaction vertex which produces the term linear in $(n_{12}\cdot n_{23}-1)$. The computation of each diagram is presented in Appendix \ref{ap:B}.} \label{3ptvertex}
\end{figure}

Next we consider the terms linear in $(n_{12}\cdot n_{23}-1)$. The diagrams that give such contributions are listed in figure \ref{3ptvertex}. The computation of each diagram is a straightforward, yet tedious task. So we relegate the detail of the computation to Appendix \ref{ap:B} and present only the final result here:  
\begin{align}
-(n_{12}\cdot n_{23}-1)\left(\frac{\lambda}{8\pi^{2}}\right)^2\left(3\zeta(3)+\frac{\pi^{2}}{3}\log\left|\frac{t_{12}t_{23}}{t_{31}\epsilon}\right|\right)\period
\end{align}

Then, after the symmetrization, we get
\beq\label{twoloop3ptfinal}
\begin{aligned}
&\langle W\!\left[\mathcal{O}_1(t_1)\mathcal{O}_2 (t_2)\mathcal{O}_3 (t_3)\right]\rangle_{\text{2-loop}} =\\
&\left(\frac{\lambda}{8\pi^{2}}\right)^2\sum_{\{i,j,k\}}\left[(n_{ij}\cdot n_{jk}-1)^{2}\left(\frac{\pi^{2}}{6}-\log\left|\frac{t_{ij}t_{jk}}{t_{ki}\epsilon}\right|+\frac{1}{2}\left(\log\left|\frac{t_{ij}t_{jk}}{t_{ki}\epsilon}\right|\right)^2\right)\right.\\
&-(n_{ij}\cdot n_{jk}-1)(n_{jk}\cdot n_{ki}-1)\left(\frac{\pi^{2}}{6}-\log\left|\frac{t_{ij}t_{jk}}{t_{ki}\epsilon}\right|\log\left|\frac{t_{jk}t_{ki}}{t_{ij}\epsilon}\right|\right)\\
&\left.-(n_{ij}\cdot n_{jk}-1)\left(3\zeta(3)+\frac{\pi^{2}}{3}\log\frac{t_{ij}t_{jk}}{t_{ki}\epsilon}\right)\right]\period
\end{aligned}
\eeq
Here again the sum is over $\{i,j,k\}=\{1,2,3\},\{2,1,3\},\{3,1,2\}$.

Comparing \eqref{twoloop3ptfinal} with the weak-coupling expansion \eqref{Bexpanded}, we can read off the two-loop anomalous dimension and the structure constant as follows: 
\begin{align}
\gamma^{(2)}_{j}&=\frac{1}{(8\pi^{2})^{2}}\left[(n_{ij}\cdot n_{jk}-1)^{2}+\frac{\pi^{2}}{3}(n_{ij}\cdot n_{jk}-1)\right]\comma\\
c_{123}^{(2)}&=\frac{1}{(8\pi^{2})^{2}}\frac{\pi^{2}}{12}\sum_{\{i,j,k\}}\left[(n_{ij}\cdot n_{jk}-1)^{2}-2(n_{ij}\cdot n_{jk}-1)(n_{jk}\cdot n_{ki}-1)\right]\period
\end{align}
As expected, the result for the anomalous dimension matches the one obtained previously in \eqref{2ptanomalous}.

\section{Three-point functions in the ladders limit\label{sec:ladder}}
In this section, we consider a special double-scaling limit called the ladders limit. The ladders limit provides an interesting solvable example of defect conformal field theories in higher dimensions: As will be explained more in detail in subsection \ref{subsec:set-up}, the only diagrams that survive in this limit are the ladder diagrams and one can sum them up by solving the Schwinger-Dyson equation. Such resummation was performed already in the literature to compute the static quark potential and the cusp anomalous dimension \cite{CHMS, ESSZ}.\footnote{In ABJ(M) theory, similar resummation of ladder diagrams in the cusp anomalous dimension was studied in \cite{ABJMladder1, ABJMladder2}.} 
A similar technique (or more precisely, a more refined version of it) can be applied to the computation of the structure constant of DCOs as we explain below.

Since this section is rather long, let us give a brief outline of what will be discussed in each subsection.
In subsection \ref{subsec:set-up}, we first explain the set up, namely the ladders limit for the two-point functions and the three-point functions. After doing so, we introduce a building block for the computation, which we call the {\it four-point kernel}, in subsection \ref{subsec:kernel} and compute it using the Schwinger-Dyson equation. We then compute the two-point function of DCOs and determine their anomalous dimensions and renormalization factors in subsection \ref{subsec:two-renorm}. 
The calculation of the three-point functions are divided into three consecutive subsections \ref{subsec:case1}-\ref{subsec:case3}, depending on the number of DCOs with non-vanishing conformal dimensions.

\subsection{Set up\label{subsec:set-up}}
\begin{figure}[t]
\centering
\includegraphics[clip,height=1.3cm]{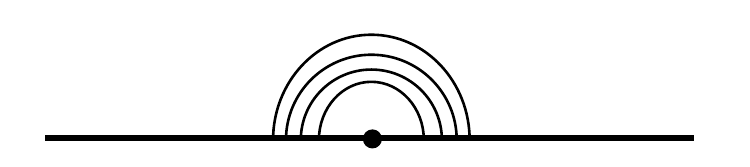}
\caption{An example of the diagrams that survive in the ladders limit.} \label{generalladder}
\end{figure}
The ladders limit was introduced in \cite{CHMS} as a double scaling limit in which the 't Hooft coupling constant $\lambda$ goes to zero while the angle between the neighboring polarizations is sent to imaginary negative infinity:
\beq
\lambda \to 0 \comma\qquad \hat{\lambda}\equiv \frac{e^{i\theta}\lambda}{4}\sim \frac{\lambda (n\cdot \tilde{n})}{2}:\text{ fixed}\comma \qquad \qquad (\cos \theta \equiv n\cdot \tilde{n})\period
\eeq
Since $\lambda$ goes to zero, all the diagrams which contain gluon propagators or interaction vertices disappear in the limit. The only diagrams that survive are the ones which have scalar propagators connecting the two segments since they come with a divergent factor $(n\cdot \tilde{n})$ which compensates the vanishing coupling constant. Such diagrams have the structure of the ladders, hence the name. (See figure \ref{generalladder}.)

Alternatively, one can define the ladders limit of DCOs directly in the zero-coupling $\mathcal{N}=4$ SYM as follows:
\beq
\begin{aligned}
  {\rm Tr}\left[{\rm Pexp}\left(  \int_{-\infty}^{t}\!\!\! d\tau \,\, \hat{\phi}_i v^{i}|\dot{x}^{\mu}|\right){\rm Pexp}\left(  \int_{t}^{\infty}\!\!\! d\tau \,\,\hat{\phi}_i \tilde{v}^{i}|\dot{x}^{\mu}|\right)\right]\period
\end{aligned}
\eeq
Here $\hat{\phi}$ is a rescaled scalar field defined by $\hat{\phi}\equiv \phi /\sqrt{\lambda}$
and $v$ and $\tilde{v}$ are complex null vectors related to the effective coupling as $\hat{\lambda}=(v\cdot \tilde{v})/2$. The equivalence of the two descriptions can be shown either at the level of diagrams or by carefully taking the limit of the polarization vectors\fn{See the discussion in section 4 of \cite{CHMS}.}. 

\begin{figure}[t]
\centering
\includegraphics[clip,height=2.5cm]{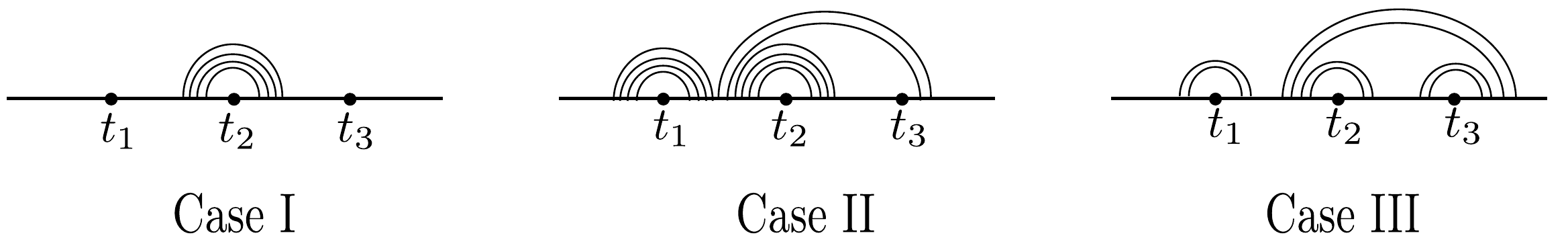}
\caption{Examples of diagrams that contribute to the Case I (left), the Case II (middle) and the Case III (right) three-point functions.} \label{casesladder}
\end{figure}

For the purpose of computing the (cusp) anomalous dimensions \cite{CHMS}, one just needs a two-point function on the Wilson loop, and we only have two polarizations to play with. On the other hand, the three-point function has three polarizations and we thus have three different angles. We can therefore define three effective couplings\fn{See \eqref{eq:defoftheta} for definitions of $\theta_i$'s}
\beq
\hat{\lambda}_i \equiv \frac{e^{i\theta_i}\lambda}{4} \qquad \qquad (i=1,2,3)\period
\eeq
and take various different limits depending on how we scale $\theta_i$'s. The simplest among them is the limit in which one of the angle, say $\theta_2$, is sent to infinity while the others are kept finite. In this limit, we have
\beq
\text{Case I}: \hat{\lambda}_1=\hat{\lambda}_3=0\comma\qquad \hat{\lambda}_2=\hat{\lambda}\neq 0\comma
\eeq
and the diagrams that survive are the ones which connect the two neighboring segments of $\mathcal{O}_2$. As we see in subsection \ref{subsec:two-renorm}, when the effective coupling is zero, the dimensions of the corresponding DCOs ($\mathcal{O}_1$ and $\mathcal{O}_3$) vanish. In what follows, we call such operators {\it trivial operators}.
The next simplest limit is the limit,
\beq
\text{Case II}: \hat{\lambda}_1\neq 0\comma\quad \hat{\lambda}_2\neq 0\comma\quad \hat{\lambda}_3=0\period
\eeq
We then have two trivial DCOs and one nontrivial DCO and the diagrams are the ones depicted in figure \ref{casesladder}.
Lastly, there is the most complicated limit, in which all three effective couplings are nonzero.
\beq
\text{Case III}: \hat{\lambda}_1\neq 0\comma\quad \hat{\lambda}_2\neq 0\comma\quad \hat{\lambda}_3\neq 0\period
\eeq
These three cases will be discussed separately in subsections \ref{subsec:case1}-\ref{subsec:case3}.

Note that the ladders limit for the three-point function can also be defined directly in the zero-coupling $\mathcal{N}=4$ SYM. For instance, the Case III corresponds to the correlator,
\beq
\begin{aligned}
&\langle {\rm Tr}\left[{\rm Pexp}\left(  \int_{-\infty}^{t_1}\!\!\! d\tau \,\,\hat{\phi}_i v_{31}^{i}|\dot{x}^{\mu}|\right){\rm Pexp}\left(  \int_{t_1}^{t_2}\!\!\! d\tau \,\, \hat{\phi}_i v_{12}^{i}|\dot{x}^{\mu}|\right)\right.\\
&\left.\quad {\rm Pexp}\left(  \int_{t_2}^{t_3}\!\!\! d\tau\,\,  \hat{\phi}_i v_{23}^{i}|\dot{x}^{\mu}|\right){\rm Pexp}\left(  \int_{t_3}^{\infty}\!\!\! d\tau \,\,\hat{\phi}_i v_{31}^{i}|\dot{x}^{\mu}|\right)\right]\rangle\comma
\end{aligned}
\eeq
where $v_{ij}$'s are complex null vectors and related to the effective couplings as $\hat{\lambda}_j=(v_{ij}\cdot v_{jk})/2$.
\subsection{Four-point kernel and the Schwinger-Dyson equation\label{subsec:kernel}}
\begin{figure}[t]
\centering
\includegraphics[clip,height=2.5cm]{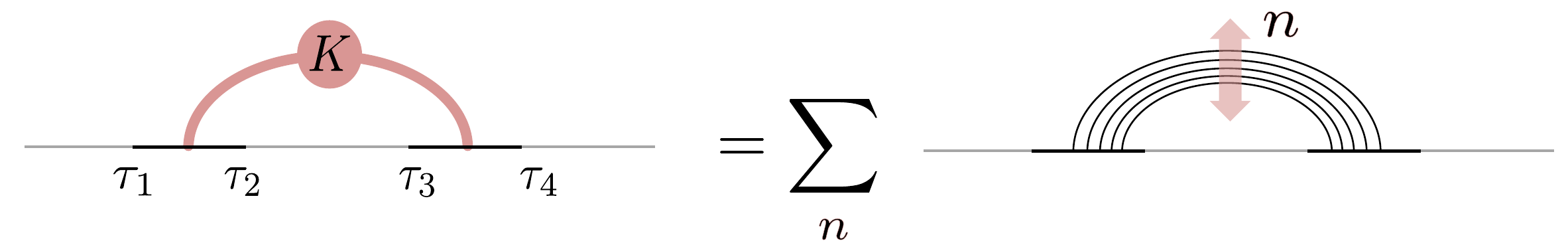}
\caption{The definition of the four-point kernel. It is defined as a sum over all the ladder diagrams with the endpoints on the black segments in the figure.} \label{defK}
\end{figure}

To compute the correlators in the ladders limit, it is useful to introduce a basic building block which we call the {\it four-point kernel} $K(\tau_1,\tau_2|\tau_3,\tau_4)$. The function $K$ is defined as a sum over all the ladder diagrams with the left endpoints in $[\tau_1,\tau_2]$ and the right endpoints in $[\tau_3,\tau_4]$. (See also figure \ref{defK})

\begin{figure}[t]
\centering
\includegraphics[clip,height=1.8cm]{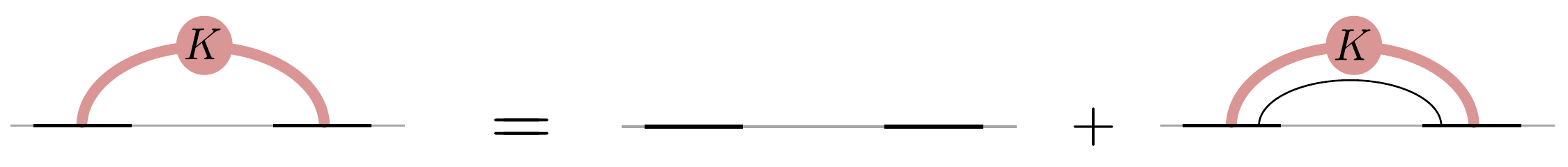}
\caption{The Schwinger-Dyson equation satisfied by the four-point kernel $K$.} \label{SDK}
\end{figure}

As shown in figure \ref{SDK}, $K(\tau_1,\tau_2|\tau_3,\tau_4)$ satisfies the Schwinger-Dyson equation
\beq\label{eq:fkernelSD}
\begin{aligned}
K(\tau_1,\tau_2|\tau_3,\tau_4)&=1+\int_{\tau_1}^{\tau_2}ds \int^{\tau_4}_{\tau_3}dt\, P(t-s)K(\tau_1,s|t,\tau_4)\comma\\
P(x)&\left(=\frac{\lambda}{8\pi^2}\frac{n\cdot \tilde{n}}{x^2}\right)=\frac{\hat{\lambda}}{4\pi^2}\frac{1}{x^2}\comma
\end{aligned}
\eeq
where $P(x)$ is a scalar propagator connecting the two segments, $[\tau_1,\tau_2]$ and $[\tau_3,\tau_4]$. By differentiating both sides, we can derive a differential equation,
\beq\label{SDdiff}
\del_{\tau_2}\del_{\tau_3} K = -P (\tau_3-\tau_2)K\period
\eeq
One can further simplify this differential equation using the conformal invariance. Owing to the conformal invariance, the kernel $K$ depends on the coordinates only through the cross ratio
\beq
u= \frac{\tau_{12}\tau_{34}}{\tau_{13}\tau_{24}}\period
\eeq
We can therefore rewrite\fn{To rewrite the differential equation, we used $\del_{\tau_2}=-\tau_{14}/(\tau_{12}\tau_{24}) \,\del_u$ and $\del_{\tau_3}=\tau_{14}/(\tau_{13}\tau_{34}) \,\del_u$} \eqref{SDdiff} as a differential equation of one variable, $u$:
\begin{align}
\left[ u(1-u)\frac{d^2}{du^2}+(1-u)\frac{d}{du}-\frac{\hat{\lambda}}{4\pi^2 (1-u)} \right] K(u)=0 \period
\end{align}
Since this is a second-order differential equation, there are two linearly independent solutions\fn{The other (incorrect) solution is $(1-u)^{-\Omega}\left[ {}_2 F_{1}^{\ast} (-\Omega,-\Omega,1;u)+\log u\,\,{}_2 F_{1} (-\Omega,-\Omega,1;u)\right]$, with ${}_2 F_{1}^{\ast} (a,b,c;u)=(\del_a+\del_b+2\del_c){}_2 F_{1}^{\ast} (a,b,c;u)$.}. The correct solution can be selected by imposing the boundary condition, $K(u=0)=1$ (or equivalently $K|_{\tau_1\to \tau_2}=K|_{\tau_3\to \tau_4}=1$), which comes from the original integral equation \eqref{eq:fkernelSD}. As a result we have
\beq
\begin{aligned}
K(u)= (1-u)^{-\Omega} {}_2 F_{1} (-\Omega,-\Omega,1;u)\comma
\end{aligned}
\eeq
with
\beq
\Omega= \frac{1}{2}\left( -1+ \sqrt{1+\frac{\hat{\lambda}}{\pi^2}} \right)\period \label{omega}
\eeq

\begin{figure}[t]
\centering
\includegraphics[clip,height=2.5cm]{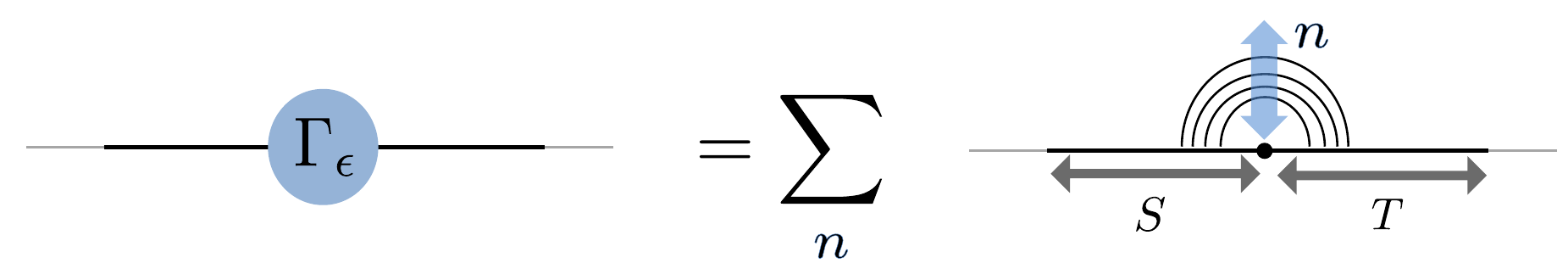}
\caption{The definition of the (regularized) vertex function. It is given by a sum of diagrams whose end-points are in $[-S,-\epsilon/2]$ and $[\epsilon/2,T]$} \label{defGamma}
\end{figure}

Using the four-point kernel, one can compute other physically important quantities. One such quantity is the {\it vertex function} $\Gamma_{\epsilon}(S,T)$, depicted in figure \ref{defGamma}. Roughly speaking, it is given by a sum over the ladder diagrams whose end points are in $[-S,0]$ and $[0,T]$. This quantity is however UV divergent and one has to introduce a point-splitting cut off to regularize such divergence. Thus, the precise definition is given by
\beq\label{defgammepsi}
\Gamma_{\epsilon} (S,T)\equiv K(-S,-\epsilon/2\,\,|\,\,\epsilon/2,T)=K\left(\frac{(S-\frac{\epsilon}{2})(T-\frac{\epsilon}{2})}{(S+\frac{\epsilon}{2})(T+\frac{\epsilon}{2})}\right)
\eeq
Here the subscript $\epsilon$ is introduced to remember that it is a ``bare" quantity which depends explicitly on the cut off. Note that the vertex function also satisfies the differential equation
\beq
\del_S \del_T\Gamma_{\epsilon}(S,T)=P(S+T) \Gamma_{\epsilon}(S,T) \label{eq:diff_eq2} \comma
\eeq
From the explicit form of the four-point kernel, one can determine the leading behavior of $\Gamma_{\epsilon}$ in the limit $S,T\gg\epsilon$,
\beq
\begin{aligned}
\Gamma_{\epsilon}(S,T)&=\underbrace{\frac{A (\Omega)}{\epsilon^{\Omega}}\left(\frac{1}{S}+\frac{1}{T}\right)^{-\Omega}}_{\equiv \Gamma_{\rm IR}(S,T)}+O(\epsilon) \comma\qquad A(\Omega)=\frac{\Gamma (2\Omega+1)}{\Gamma (\Omega+1)^2}\period
\end{aligned}
\eeq

As discussed in Appendix \ref{ap:sch}, there is an intriguing relation between the vertex function and the solutions to the Schr\"{o}dinger equation (also called the Bethe-Salpeter equation) studied in \cite{ESSZ,CHMS}: By rewriting it in a different coordinate, one can explicitly show that the leading term $\Gamma_{\rm IR}$ is related to the ground-state wave function of the Schr\"{o}dinger equation. Furthermore, it turns out that the subleading terms in the expansion correspond to the excited-state wave functions of the same Schr\"{o}dinger equation. In this sense, the full vertex function $\Gamma_{\epsilon}$ can be regarded as a generating function of such wave functions. From the defect CFT point of view,  
the expansion of $\Gamma_{\epsilon}$ in $\epsilon$ is nothing but the OPE expansion of the ``four-point function" $K$, with its leading term controlled by a conformal primary DCO and the subleading terms controlled by its descendants. This provides a physical interpretation of the solutions to the Schr\"{o}dinger equation: Namely the ground state corresponds to a conformal primary and the excited states correspond to its descendants.  For a more detailed account on this point, see Appendix \ref{ap:sch}.

\subsection{Two-point functions and renormalization\label{subsec:two-renorm}}
As in the two-loop analysis performed in the preceding sections, to extract the scheme independent structure constant, one first needs to determine the renormalization factor $Z^{-1/2}$:
\beq
\mathcal{O}=Z^{-1/2}\mathcal{O}^{B}\comma 
\eeq
Here $\mathcal{O}^{B}$ denotes the bare DCO while $\mathcal{O}$ denotes the renormalized DCO. The renormalization factor is determined by requiring that the renormalized two-point function has a canonical form:
\begin{align}
\langle \mathcal{O}(\tau_1) \mathcal{O}(\tau_2) \rangle = Z^{-1} \langle \mathcal{O}^B(\tau_1) \mathcal{O}^B(\tau_2) \rangle = \frac{1}{|\tau_{12}|^{2\Delta}} \period \label{eq:two-point}
\end{align}
In what follows, we evaluate the two-point function of the bare operators and determine $Z$.

\begin{figure}[t]
\centering
\includegraphics[clip,height=5cm]{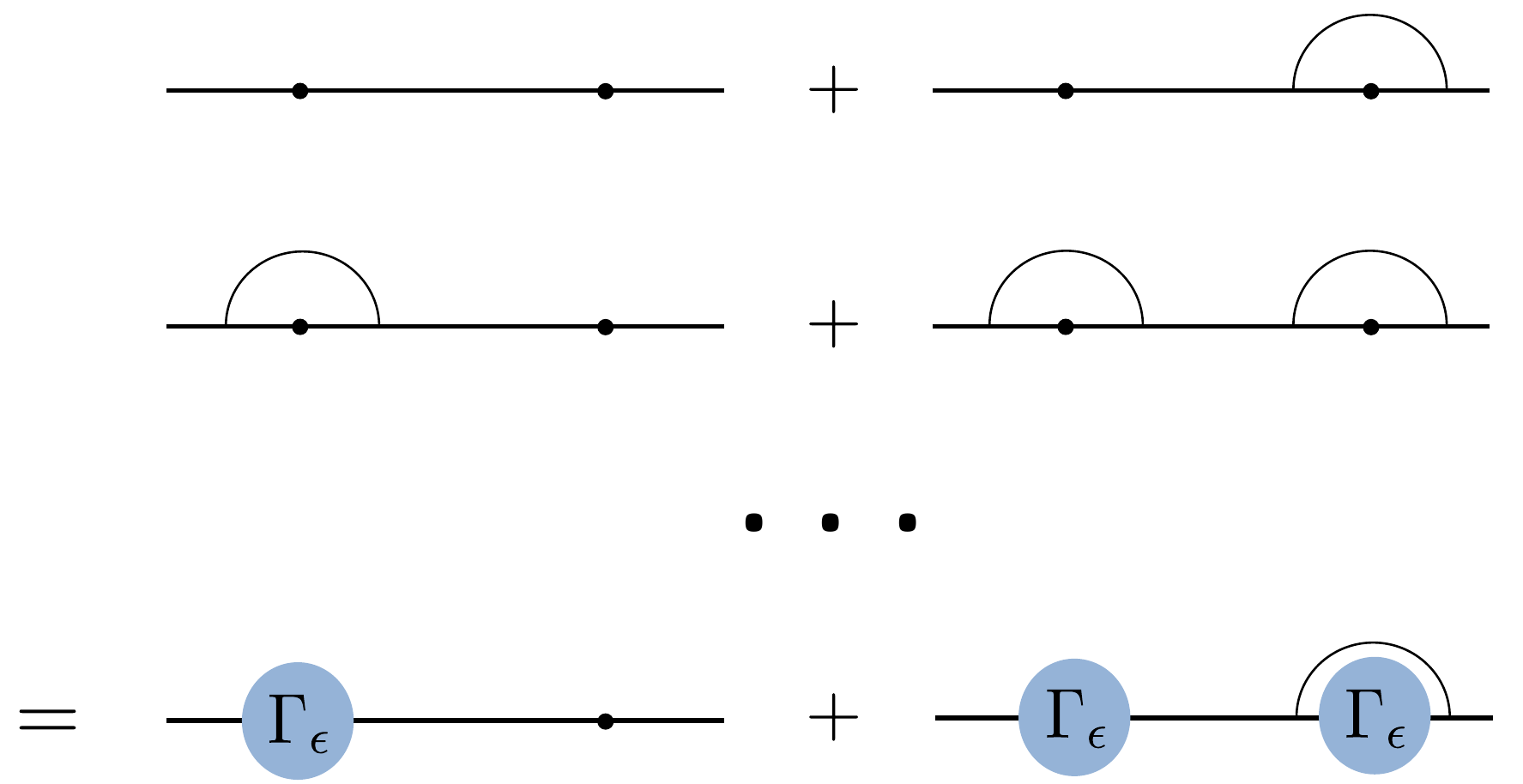}
\caption{The resummation of the ladder diagrams for the two-point function. The contribution from the left figure on the last line will be denoted by ${\tt 1st}$ while the one from the right figure will be denoted as ${\tt 2nd}$.} \label{resum2pt}
\end{figure}
The two-point function can be computed using the vertex functions as shown in figure \ref{resum2pt}. The contribution from each diagram is given as follows:
\beq
\begin{aligned}
{\tt 1st} &= \Gamma_{\epsilon}(\infty,\tau_{21})\comma\\
{\tt 2nd}&= \int^{\tau_2^{-}}_{\tau_1^{+}}ds \int^{\infty}_{\tau_2^{+}}dt\, \Gamma_{\epsilon}(\infty,s-\tau_1)P(t-s)\Gamma_{\epsilon}(\tau_2-s,t-\tau_2)\period
\end{aligned}
\eeq
As in the preceding sections, the endpoints are regularized as $\tau_i^{\pm}=\tau_i \pm \epsilon/2$.
The first contribution can be approximated by $\Gamma_{\rm IR}$ and is given by
\beq
{\tt 1st} \sim \left(\frac{\tau_{21}}{\epsilon}\right)^{\Omega}\period
\eeq
To evaluate the second contribution, we use the differential equation \eqref{eq:diff_eq2} and rewrite it as
\begin{align}\label{eq:two-second}
{\tt 2nd}&=\int_{\tau_1^{+}}^{\tau_2^{-}} ds \int_{\tau_2^{+}}^{\infty} dt \,\Gamma_{\epsilon}(\infty, s-\tau_1) (-\partial_s\partial_t) \Gamma_{\epsilon} (\tau_2-s, t-\tau_2) \\
&= -\int_{\tau_1^{+}}^{\tau_2^{-}} ds \, \Gamma_{\epsilon}(\infty, s-\tau_1) \partial_s \Gamma_{\epsilon} (\tau_2-s, \infty)
\end{align}
In passing to the second line, we used $\Gamma _{\epsilon} (\ast,\epsilon/2)=K (u)|_{u=0}=0$. 
Unlike the first contribution, there is a priori no reason to expect that $\Gamma_{\epsilon}$ can be approximated by $\Gamma_{\rm IR}$ in \eqref{eq:two-second} since the arguments of $\Gamma_{\epsilon}$ can be of order $O(\epsilon)$. Nevertheless, it turns out that the leading singular piece in the limit $\epsilon\ll 1$ can be computed by replacing $\Gamma_{\epsilon}$ with $\Gamma_{\rm IR}$. Roughly speaking, this is because the difference between $\Gamma_{\epsilon}$ and $\Gamma_{\rm IR}$,
\beq
\Gamma_{\rm UV}\equiv\Gamma_{\epsilon}-\Gamma_{\rm IR} \comma
\eeq 
is of order $\epsilon \,\Gamma_{\rm IR} $ whenever the arguments are $O(1)$ while it is of $O(1)$ only when the arguments are in a small interval of length $\epsilon$ near the origin. Therefore the contribution from $\Gamma_{\rm UV}$ is always $O(\epsilon)$ smaller than the contribution from $\Gamma_{\rm IR}$. See Appendix \ref{ap:gamUV} for more detailed arguments.

Once we replace $\Gamma_{\epsilon}$ with $\Gamma_{\rm IR}$, the computation is straightforward:
\beq
\begin{aligned}
{\tt 2nd}&\overset{\epsilon\ll 1}{=} -\left(\frac{A(\Omega)}{\epsilon^{\Omega}}\right)^2\int_{\tau_1}^{\tau_2} ds \, (s-\tau_1)^{\Omega} \partial_s  (\tau_2-s)^{\Omega}=\left(\frac{A(\Omega)}{\epsilon^{\Omega}}\right)^2\Omega\int_{0}^{1} d\bar{s} \, \bar{s}^{\Omega} (1-\bar{s})^{\Omega-1}\\
&\,=\frac{\Gamma (2\Omega+1)}{\Gamma (\Omega +1)^2} \frac{|\tau_{12}|^{2\Omega}}{\epsilon^{2\Omega}}\period
\end{aligned}
\eeq
In the second equality, we performed the transformation $\bar{s}=(s-\tau_1)/(\tau_1-\tau_2)$ and in the last equality we used
\beq\label{integralIab}
I(a,b)\equiv b \int_{0}^{1} d\bar{s} \, \bar{s}^{a} (1-\bar{s})^{b-1}=\frac{\Gamma (a+1)\Gamma(b+1)}{\Gamma (a+b+1)}\period
\eeq
We thus obtain
\beq\label{baredcotwo}
\langle \mathcal{O}^{B}(\tau_1)\mathcal{O}^{B}(\tau_2)\rangle \overset{\epsilon\ll 1}{=}\frac{\Gamma (2\Omega+1)}{\Gamma (\Omega +1)^2} \frac{|\tau_{12}|^{2\Omega}}{\epsilon^{2\Omega}}\period
\eeq
Note that the first contribution ${\tt 1st}$ only gives a subleading correction. By comparing \eqref{baredcotwo} with \eqref{eq:two-point}, we conclude that the conformal dimension and the renormalization factor of the DCO are given by\fn{Note that the conformal dimension of the DCO is negative. This however is not a contradiction since the defect CFT we are studying lacks unitarity.}
\beq
\Delta (\hat{\lambda}) = -\Omega\comma \qquad Z (\hat{\lambda})= \frac{A (\Omega)}{\epsilon^{2\Omega}}\period
\eeq  
As expected, the result for the conformal dimension matches the one in \cite{CHMS}.

Before ending this subsection, let us make a further comment on $\Gamma_{\epsilon}$ and $\Gamma_{\rm IR}$. As explained above, the leading term in $\epsilon\ll 1$ limit can be computed by replacing $\Gamma_{\epsilon}$'s with $\Gamma_{\rm IR}$'s in the diagram with the maximal number of vertex functions. As discussed in Appendix \ref{ap:gamUV}, this is rather an universal phenomenon and it applies also to the computation of the multi-point functions. When the effect of the renormalization is taken into account, this leads to the following recipe of computing the {\it renormalized} correlation functions:
\begin{enumerate}
\item List up all the diagrams that compute the bare correlators and select those with a maximal number of vertex functions.
\item Replace $\Gamma_{\epsilon}$ in those diagrams with the renormalized vertex function $\Gamma^{R}$
\beq
\Gamma^{R} (S,T) \equiv \lim_{\epsilon \rightarrow 0} Z^{-1/2} \Gamma_{\rm IR}(S,T) = \sqrt{A(\Omega)} \left( \frac{1}{S}+\frac{1}{T} \right)^{-\Omega} \comma \label{eq:Gamma_R}
\eeq
and compute the integrals.
\end{enumerate}
In the subsections below, we compute the structure constants following this recipe.
\subsection{Case I: 1 nontrivial and 2 trivial DCOs\label{subsec:case1}}
\begin{figure}[t]
\centering
\includegraphics[clip,height=1.5cm]{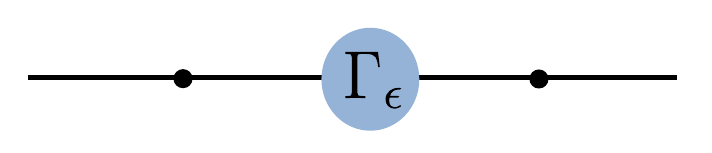}
\caption{The resummation of the ladder diagrams for the Case I three-point function.} \label{case11}
\end{figure}
We now compute the three-point function of one nontrivial DCO and two trivial DCOs. For simplicity we first consider the case where the operator in the middle is a nontrivial DCO (see figure \ref{case11}). In this case, one can easily resum the diagrams by a single vertex function. Thus, following the prescription at the end of the previous subsection, we obtain
\beq
\begin{aligned}
\langle\mathcal{O}_1^{\circ}(\tau_1)\mathcal{O}_2^{\bullet}(\tau_2)\mathcal{O}_3^{\circ}(\tau_3) \rangle&=\Gamma^{R}(\tau_{21},\tau_{32})=\frac{C_{\circ \bullet \circ}}{\tau_{21}^{-\Omega}\tau_{32}^{-\Omega}\tau_{31}^{\Omega}}\comma\\
C_{\circ\bullet\circ}&=\sqrt{A(\Omega)}=\frac{\Gamma (2\Omega+1)^{1/2}}{\Gamma (\Omega+1)}\period
\end{aligned}
\eeq
Here and in what follows, the symbols $\circ$ and $\bullet$ signify a trivial DCO and a nontrivial DCO respectively. At weak coupling, the result can be expanded from (\ref{omega}) as
\beq
C_{\circ \bullet \circ} = 1+ \frac{\pi^2}{12} \left( \frac{\hat{\lambda}}{4\pi^2} \right)^2 -\left(\frac{\pi^2}{6}+\zeta(3) \right) \left( \frac{\hat{\lambda}}{4\pi^2} \right)^3 +O(\hat{\lambda}^4) \period  \label{eq:weak_1}
\eeq
The result up to two loops reproduces the perturbative result for the ladder diagrams in the previous section. At strong coupling $(\hat{\lambda}\gg 1)$, on the other hand, the result exponentiates and is given by
\beq
\ln C_{\circ \bullet \circ}\sim \frac{\sqrt{\hat{\lambda}}}{2\pi}\ln 2\period
\eeq
The structure of the result suggests the existence of some dual description in terms of classical string with the tension $\sqrt{\hat{\lambda}}$. It would be interesting to look for such a description which reproduces this result.

\begin{figure}[t]
\centering
\includegraphics[clip,height=2cm]{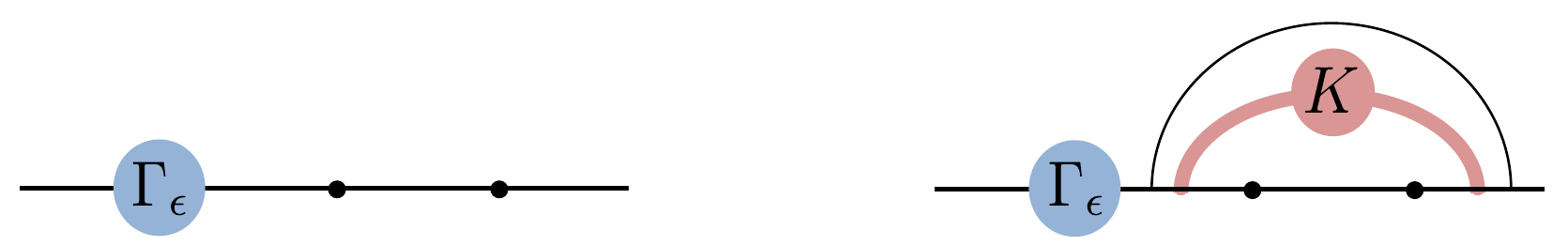}
\caption{The resummation of the ladder diagrams for the Case I three-point function when the leftmost operator is the nontrivial DCO. Unlike the one in figure \ref{case11}, the result is given by a sum of two diagrams shown above (although the structure constant itself must be the same).} \label{case12}
\end{figure}

As a consistency check, let us also compute the three-point function in which the rightmost operator is a nontrivial DCO $(C_{\bullet\circ\circ})$. Although the result must be equal to $C_{\circ \bullet\circ}$ owing to the permutation symmetry, the diagrams are not identical as shown in figure \ref{case12}. Written explicitly, we have
\beq\label{leftbullet}
\frac{C_{\bullet\circ\circ}}{\tau_{21}^{-\Omega}\tau_{32}^{\Omega}\tau_{31}^{-\Omega}}=\Gamma^{R}(\infty, \tau_{21})+\underbrace{\frac{\hat{\lambda}}{4\pi^2}\int_{\tau_1}^{\tau_2}ds\int^{\infty}_{\tau_3}dt\, \frac{\Gamma^{R}(\infty,s-\tau_1)K(s,\tau_2|\tau_3,t)}{(t-s)^2}}_{(\ast)}\period
\eeq
Using the Schwinger-Dyson equation \eqref{SDdiff}, one can rewrite the second term on the right hand side as
\beq
(\ast)=-\int_{\tau_1}^{\tau_2}ds\int^{\infty}_{\tau_3}dt\, \Gamma^{R}(\infty,s-\tau_1)\del_s\del_t K(s,\tau_2|\tau_3,t)\period
\eeq
Clearly, the $t$ integral can be trivially performed and we get
\beq
(\ast)=\int^{\tau_2}_{\tau_1}ds \,\del_s \left(1-K(s,\tau_2|\tau_3,\infty) \right)\Gamma^{R}(\infty,s-\tau_1)\comma
\eeq
where we used $K(\ast,\ast^{\prime}|x,x)=1$. By performing the integration by parts and adding the first term in \eqref{leftbullet}, we arrive at a simple formula
\beq\label{simpleonenontrivial}
\frac{C_{\bullet\circ\circ}}{\tau_{21}^{-\Omega}\tau_{32}^{\Omega}\tau_{31}^{-\Omega}}=\int^{\tau_2}_{\tau_1}ds\, \del_s \Gamma^{R}(\infty,s-\tau_1)\, K(s,\tau_2|\tau_3,\infty)\period
\eeq
As shown in Appendix \ref{ap:integral}, this integral can be evaluated explicitly using the identities of the hypergeometric functions and we obtain the expected result
\beq
C_{\bullet\circ\circ}=C_{\circ\bullet\circ}=\sqrt{A(\Omega)}\period
\eeq

\subsection{Case II: 2 nontrivial and 1 trivial DCOs\label{subsec:case2}}
Let us now compute the structure constant of the case II correlators,
\beq
\langle \mathcal{O}_1^{\bullet}(\tau_1)\mathcal{O}_2^{\bullet}(\tau_2)\mathcal{O}_3^{\circ}(\tau_3)\rangle=\frac{C_{\bullet\bullet\circ}}{\tau_{21}^{-\Omega_1-\Omega_2}\tau_{32}^{\Omega_1-\Omega_2}\tau_{31}^{-\Omega_1+\Omega_2}} \period
\eeq
Since we have two effective couplings $\hat{\lambda}_{1,2}$ in this case, we shall distinguish various functions of the effective couplings by putting subscripts; for instance the dimension of each operator will be denoted as $\Delta_i=-\Omega_i\equiv -\Omega (\hat{\lambda}_i)$. 

\begin{figure}[t]
\centering
\includegraphics[clip,height=4cm]{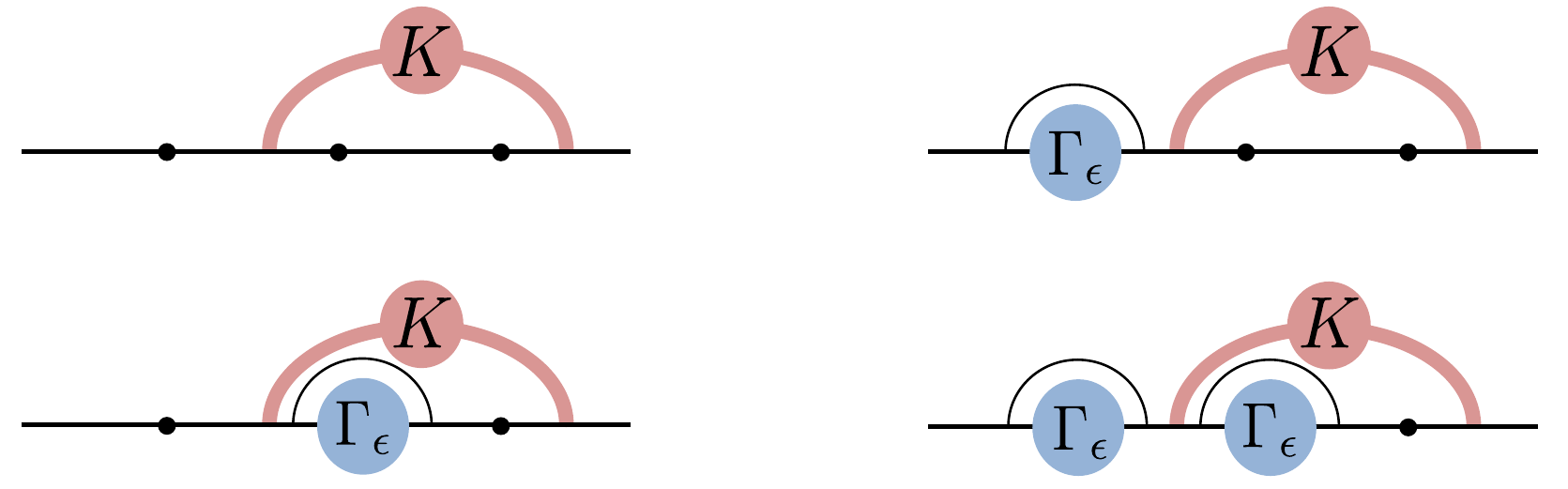}
\caption{The resummation of the ladder diagrams for the Case II three-point function. Among the four diagrams, only the last one gives a dominant contribution in the limit $\epsilon\to 0$.} \label{case2}
\end{figure}
The diagrams that contribute to the bare three-point functions are listed in figure \ref{case2}. Following the recipe in subsection \ref{subsec:two-renorm}, we just consider the diagram with a maximal number of vertex functions, which in this case is the last diagram. We then get the following expression for the renormalized three-point function\fn{As mentioned above, the notations $\Gamma_i^{R}$ and $K_i$ mean $\Gamma^{R}(\hat{\lambda}_i)$ and $K(\hat{\lambda}_i)$.}
\beq\label{twonontrivialintegral}
\begin{aligned}
\langle \mathcal{O}_1^{\bullet}\mathcal{O}_2^{\bullet}\mathcal{O}_3^{\circ}\rangle&=\int^{\tau_2}_{\tau_1}ds\int^{\tau_3}_{\tau_2}dt \int^{\tau_1}_{-\infty}du\int^{s}_{\tau_1}dv \,\,(-\del_u\del_v)\Gamma_1^{R}(\tau_1-u,v-\tau_1)\\
&\qquad\qquad \qquad \qquad \times K_1(v,s|\tau_3,\infty)(-\del_s \del_t)\Gamma_2^{R}(\tau_2-s,t-\tau_2)\period 
\end{aligned}
\eeq 
Here we have already replaced the scalar propagators with the derivatives using the equation \eqref{SDdiff}. In \eqref{twonontrivialintegral}, the integral of $u$ can be performed straightforwardly. After doing so, the integral of $v$ becomes
\beq
\int^{s}_{\tau_1} dv\, \del_v \Gamma_1^{R}(\infty,v-\tau_1)K_1(v,s|\tau_3,\infty)\period
\eeq
This integral is essentially the same as the right hand side of \eqref{simpleonenontrivial}. Therefore, using the result there, we can evaluate it to get
\begin{align}
\langle \mathcal{O}_1^{\bullet}\mathcal{O}_2^{\bullet}\mathcal{O}_3^{\circ}\rangle&=\tau_{31}^{\Omega_1}\sqrt{A(\Omega_1)}\int^{\tau_2}_{\tau_1}ds\int^{\tau_3}_{\tau_2}dt \left(\frac{s-\tau_1}{\tau_3-s}\right)^{\Omega_1}(-\del_s \del_t)\Gamma_2^{R}(\tau_2-s,t-\tau_2)\\
&=\tau_{31}^{\Omega_1}\sqrt{A(\Omega_1)A(\Omega_2)}\int^{\tau_2}_{\tau_1}ds\int^{\tau_3}_{\tau_2}dt \left(\frac{s-\tau_1}{\tau_3-s}\right)^{\Omega_1}(-\del_s \del_t)\left(\frac{(\tau_2-s)(t-\tau_2)}{t-s}\right)^{\Omega_2}\period\nn
\end{align}
We can now perform the $t$ integral to get
\beq
\langle \mathcal{O}_1^{\bullet}\mathcal{O}_2^{\bullet}\mathcal{O}_3^{\circ}\rangle= -\tau_{31}^{\Omega_1}\tau_{32}^{\Omega_2}\sqrt{A(\Omega_1)A(\Omega_2)}\int^{\tau_2}_{\tau_1}ds \left(\frac{s-\tau_1}{\tau_3-s}\right)^{\Omega_1}\del_s \left(\frac{\tau_2-s}{\tau_3-s}\right)^{\Omega_2}\period
\eeq
The last integral can be done explicitly\fn{The integral reduces to the integral for $I(a,b)$ given in \eqref{integralIab}.} by performing the following change of variables, which amounts to performing the M\"{o}bius transformation  $(\tau_1,\tau_2,\tau_3)\to (0,1,\infty)$:
\beq
\bar{s} =\frac{s-\tau_1}{s-\tau_3}\frac{\tau_2-\tau_3}{\tau_2-\tau_1}\period
\eeq
As a result, we get
\beq
\begin{aligned}
C_{\bullet\bullet\circ}=\sqrt{A(\Omega_1)A(\Omega_2)}I(\Omega_1,\Omega_2)=\frac{\Gamma (2\Omega_1+1)^{1/2}\Gamma (2\Omega_2+1)^{1/2}}{\Gamma (\Omega_1+\Omega_2+1)}\period
\end{aligned}
\eeq
Note that the result (correctly) reduces to the one in the previous subsection if we set $\Omega_2=0$.
At weak coupling, the result reads
\beq
C_{\bullet\bullet\circ}=1+\frac{(\hat{\lambda}_1-\hat{\lambda}_2)^2}{192\pi^2}+\frac{(\hat{\lambda}_1-\hat{\lambda}_2)^2(\hat{\lambda}_1+\hat{\lambda}_2)(\pi^2+6\zeta(3))}{384\pi^6}+O(\hat{\lambda}^4)\comma
\eeq
whereas at strong coupling it reads
\beq
\ln C_{\bullet\bullet\circ}\sim \Omega_1 \ln \frac{2\Omega_1}{\Omega_1+\Omega_2}+\Omega_2 \ln \frac{2\Omega_2}{\Omega_1+\Omega_2} \comma\qquad \qquad \Omega_i \sim \frac{\sqrt{\hat{\lambda}_i}}{2\pi}\period
\eeq
The result at two loops at weak coupling matches the ladder contribution to the perturbative result given in section \ref{sec:three-point}.
\subsection{Case III: 3 nontrivial DCOs\label{subsec:case3}}
\begin{figure}[t]
\centering
\includegraphics[clip,height=12cm]{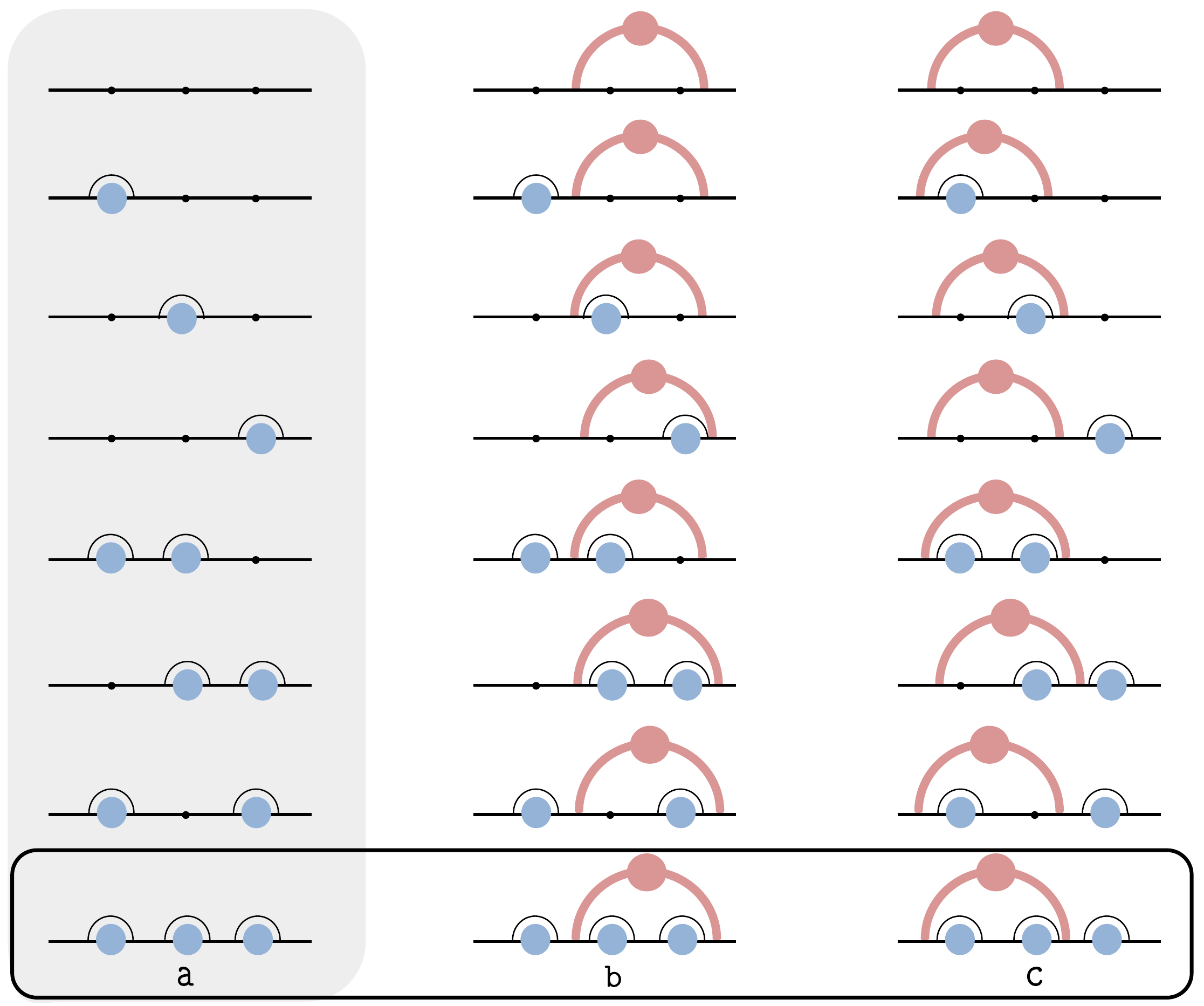}
\caption{The resummation of the ladder diagrams for the Case III three-point function. To avoid the double-counting, the diagrams in the first column (depicted in the gray background) must be summed with a negative sign. Among these diagrams only the last three (encircled by the thick black line) contribute dominantly in the limit $\epsilon\to 0$.} \label{case3}
\end{figure}

In this subsection, we compute the most general three-point functions of DCOs,
\beq
\langle\mathcal{O}_1^{\bullet}(\tau_1)\mathcal{O}_2^{\bullet}(\tau_2)\mathcal{O}_3^{\bullet}(\tau_3) \rangle=\frac{C_{\bullet\bullet\bullet}}{\tau_{21}^{-\Omega_1-\Omega_2+\Omega_3}\tau_{32}^{-\Omega_2-\Omega_3+\Omega_1}\tau_{31}^{-\Omega_3-\Omega_1+\Omega_2}}\period
\eeq
As shown in figure \ref{case3}, there are several different diagrams that contribute to this three-point function. However, only three diagrams ({\tt a}, {\tt b} and {\tt c} in figure \ref{case3}) contain a maximal number of vertex functions. Below we compute the contributions from these diagrams following the recipe in subsection \ref{subsec:two-renorm}.

The contribution from the diagram ${\tt a}$ is given by
\beq
\begin{aligned}
{\tt a}=(-1)\times \int^{\tau_2}_{\tau_1}ds \int^{\tau_3}_{\tau_2}dt \int^{\tau_3}_{t} du \int^{\infty}_{\tau_3}dv \, \Gamma^{R}(\infty,s-\tau_1)(-\del_s\del_t)\Gamma^{R}(\tau_2-s,t-\tau_2)&\\
\times  (-\del_u\del_v)\Gamma^{R}(\tau_3-u,v-\tau_3)\period&
\end{aligned}
\eeq
Here again we have already replaced the scalar propagators with the derivatives using \eqref{SDdiff}. Since the integrand is a total derivative with respect to $u$ and $v$, one can perform those integrals to get
\beq
\begin{aligned}
&{\tt a}=-I_1\comma\\
&I_1\equiv\prod_{i=1}^{3}\sqrt{A(\Omega_i)}\int_{\tau_1}^{\tau_2}ds \int_{\tau_2}^{\tau_3}dt \, (s-t)^{\Omega_1}(-\del_s\del_t)\left(\frac{(\tau_2-s)(t-\tau_2)}{t-s}\right)^{\Omega_2} (\tau_3-t)^{\Omega_3} \period
\end{aligned}
\eeq

Next we consider the contribution from the diagram {\tt b}, which is given by
\beq
\begin{aligned}
{\tt b}=&\int^{\tau_2}_{\tau_1} ds \int^{\tau_3}_{\tau_2}dt \int^{\infty}_{\tau_3}du\int^{\tau_3}_{t}dv\int^{\tau_1}_{-\infty}dw \int^{s}_{\tau_1} dx\, K(x,s|u,\infty)\\
&\times (-\del_w \del_x)\Gamma^{R}(w-\tau_1,x-\tau_1)(-\del_s \del_t)\Gamma^{R}(\tau_2-s,t-\tau_2)(-\del_u\del_v)\Gamma^{R}(\tau_3-v,u-\tau_3)\period
\end{aligned}
\eeq
By performing the trivial integrals of $v$ and $w$, we get the integrand with $\del_u \Gamma^{R}(\infty,u-\tau_3)$ and $\del_x \Gamma^{R}(\infty,x-\tau_1)$. Then, the integral of $x$ coincides with the one we already studied in \eqref{simpleonenontrivial}. We can thus evaluate it to obtain
\beq
\begin{aligned}
{\tt b}=\prod_{i=1}^{3}\sqrt{A (\Omega_i)}\int^{\tau_2}_{\tau_1} ds \int^{\tau_3}_{\tau_2}dt \int^{\infty}_{\tau_3}du\left(\frac{(s-\tau_1)(u-\tau_1)}{u-s}\right)^{\Omega_1}&\\
\times (-\del_s\del_t)\left(\frac{(\tau_2-s)(t-\tau_2)}{t-s}\right)^{\Omega_2}\del_u \left(\frac{(\tau_3-t)(u-\tau_3)}{u-t}\right)^{\Omega_3}\period&
\end{aligned}
\eeq
After doing so, we perform integration by parts for the $u$ integral. We then get $I_1$ as a surface term and the full result reads
\beq
{\tt b}=I_1 + I_2\comma
\eeq
with
\beq
\begin{aligned}
I_2=\prod_{i=1}^{3}\sqrt{A (\Omega_i)}\int^{\tau_2}_{\tau_1} ds \int^{\tau_3}_{\tau_2}dt \int^{\infty}_{\tau_3}du\,(-\del_u)\left(\frac{(s-\tau_1)(u-\tau_1)}{u-s}\right)^{\Omega_1}&\\
\times (-\del_s\del_t)\left(\frac{(\tau_2-s)(t-\tau_2)}{t-s}\right)^{\Omega_2}\, \left(\frac{(\tau_3-t)(u-\tau_3)}{u-t}\right)^{\Omega_3}\period
\end{aligned}
\eeq

The contribution from the diagram {\tt c} can be evaluated in a similar manner and the result reads
\beq
\begin{aligned}
{\tt c}=I_3\equiv\prod_{i=1}^{3}\sqrt{A (\Omega_i)}\int^{\tau_2}_{\tau_1} ds \int^{\tau_3}_{\tau_2}dt \int^{\tau_1}_{-\infty}du\,(-\del_u)\left(\frac{(s-\tau_1)(u-\tau_1)}{u-s}\right)^{\Omega_1}&\\
\times (-\del_s\del_t)\left(\frac{(\tau_2-s)(t-\tau_2)}{t-s}\right)^{\Omega_2}\, \left(\frac{(\tau_3-t)(u-\tau_3)}{u-t}\right)^{\Omega_3}\period
\end{aligned}
\eeq
Summing up three contributions, we get
\beq
\langle \mathcal{O}_1^{\bullet}\mathcal{O}_2^{\bullet}\mathcal{O}_3^{\bullet}\rangle={\tt a}+{\tt b}+{\tt c}=I_2+I_3
\eeq
To further proceed, we perform the following change of variables,
\beq
\bar{s}=\frac{\tau_{23}}{\tau_{21}}\frac{s-\tau_1}{s-\tau_3}\comma\qquad \bar{t}= \frac{\tau_{31}}{\tau_{32}}\frac{t-\tau_2}{t-\tau_1}\comma\qquad \bar{u}=\frac{\tau_{12}}{\tau_{13}}\frac{u-\tau_3}{u-\tau_2}\period
\eeq
After this change of variables, the spacetime dependence of the three-point function comes out naturally as a factorized prefactor, and the rest can be combined into a single integral,
\beq\label{case3integral}
\begin{aligned}
C_{\bullet\bullet\bullet}&=\left(\prod_{i=1}^{3}\sqrt{A (\Omega_i)}\right)\times J\comma \\
J&\equiv -  \int_{0}^{1}d\bar{s}\int_{0}^{1}d\bar{t}\int_{0}^{1}d\bar{u}\,\del_{\bar{u}} \left[f(\bar{u},\bar{s})^{-\Omega_1}\right]\del_{\bar{s}}\left[f(\bar{s},\bar{t})^{-\Omega_2}\right]\del_{\bar{t}}\left[f(\bar{t},\bar{u})^{-\Omega_3}\right]\period
\end{aligned}
\eeq
with $f (x,y)\equiv (1-x)^{-1}+y^{-1}-1$. The integral is clearly invariant under any permutation\fn{The invariance under the cyclic permutation can be seen by the permutation of the integrated variables while the reflection invariance can be shown by performing $x\to 1-x$ to all the integrated variables.} of $\Omega_i$'s. It is also possible to evaluate the integral $J$ explicitly and express it as an infinite sum, although the result is rather long. See Appendix \ref{ap:sum} for the explicit expression.

Let us now expand the result at weak and strong couplings. At weak coupling, one obtains
\begin{align}
\begin{split}
&C_{\bullet \bullet \bullet} = 1+ \frac{1}{192\pi^2}\left( \sum_i^3 \hat{\lambda}_i^2-2\sum_{i<j}\hat{\lambda}_i\hat{\lambda}_j \right) \\
&+\frac{\pi^2+6\zeta(3)}{384\pi^6} \left( \sum_{i=1}^3 \hat{\lambda}_i^3-\sum_{i<j} (\hat{\lambda}_i^2 \hat{\lambda}_j+\hat{\lambda}_i\hat{\lambda}_j^2) \right) +\frac{\pi^2-3-6\zeta(3)}{192\pi^6} \hat{\lambda}_1\hat{\lambda}_2\hat{\lambda}_3 +\mathcal{O}(\hat{\lambda}^4) \period
\end{split} 
\end{align} 
One can easily check that the result is consistent with the direct perturbative computation in section \ref{sec:three-point}. Also, by sending one of the effective coupling to zero, one recovers the result in the previous subsection\fn{The last term at three loops $\hat{\lambda}_1\hat{\lambda}_2\hat{\lambda}_3$ is truly a new term, which didn't show up in the results in the preceding subsections. It would be interesting to understand it from the integrability point of view.}. 

To study the leading behavior at strong coupling $(\hat{\lambda}_i\gg 1)$, we use the saddle-point approximation of the integral $J$, 
\beq
\begin{aligned}
&J= \int_{0}^{1}d\bar{s} \int^{1}_{0} d\bar{t}\int^{1}_{0}d\bar{u} \,e^{g(\bar{s},\bar{t},\bar{u})}\comma\\
&g(\bar{s},\bar{t},\bar{u})=-\Omega_1 \log f(\bar{u},\bar{s})- \Omega_2 \log f(\bar{s},\bar{t})-\Omega_3 \log f(\bar{t},\bar{u}) +O(\ln \hat{\lambda})\period
 \end{aligned}
\eeq
The subleading term $O(\ln \hat{\lambda})$ can be neglected for the computation of the leading strong-coupling behavior. The saddle-point equation $\del_{\bar{s}}g=\del_{\bar{t}}g=\del_{\bar{u}}g=0$ has two solutions, but one of them is outside the integration region $\bar{s},\bar{t},\bar{u}\in [0,1]$ and should be discarded. The one inside the integration region is given by
\beq
\left(\bar{s}^{\ast},\bar{t}^{\ast},\bar{u}^{\ast}\right)=\left(\frac{\Omega_1}{\Omega_1+\Omega_2},\frac{\Omega_2}{\Omega_2+\Omega_3},\frac{\Omega_3}{\Omega_3+\Omega_1}\right)\period
\eeq
Evaluating $g$ at this saddle point and multiplying the prefactors $\prod_i \sqrt{A(\Omega_i)}$, we obtain
\begin{align}
\ln C_{\bullet\bullet\bullet}\sim\Omega_{1}\log\frac{2\Omega_{1}}{\Omega_{1}+\Omega_{2}+\Omega_{3}}+\Omega_{2}\log\frac{2\Omega_{2}}{\Omega_{1}+\Omega_{2}+\Omega_{3}}+\Omega_{3}\log\frac{2\Omega_{3}}{\Omega_{1}+\Omega_{2}+\Omega_{3}} \period \label{eq:strong_3}
\end{align} 
Again the structure of the result suggests that there should be some interpretation in terms of classical string, and it would be extremely interesting if we can re-derive this result from such a perspective.

\section{Discussion\label{sec:discussion}}
In this paper, we computed the structure constants on the $1/2$ BPS Wilson loop at weak coupling. We first performed the computation at two loops and then resummed the diagrams in the ladders limit. In what follows, we indicate a few possible future directions.

\subsubsection*{Relation to integrability}
As mentioned in the introduction, one of our main motivations was to provide data for the integrability-based approach. In this regard, a natural next step is to reproduce the results in this paper by generalizing the {\it hexagon approach}\fn{In a recent paper \cite{treelevel3pt}, tree-level three-point functions of non-BPS operators on the Wilson loop were computed using integrability and the open-string version of the hexagon approach was proposed.} proposed in \cite{BKV}. Preliminary investigation suggests that the process involving four (mirror) particles starts to contribute at as early as two loops while the process involving six particles shows up at three loops. Such processes appear much later in the ordinary three-point functions\fn{At 10 and 14 loops respectively.}, and  it is a great advantage of studying DCO's that we can explore them with the results already available. 

From the string worldsheet point of view, the DCO's studied in this paper correspond to the {\it boundary condition changing operators}. This viewpoint might provide an alternative approach to the three-point function on the Wilson loop which is based on the form factor expansion of the boundary condition changing operators. See \cite{BH} for recent discussions on such form factors in integrable field theories.
\subsubsection*{Ladders limit}
The ladders limit of $\mathcal{N}=4$ SYM provides an invaluable example of the (defect) conformal field theories in higher dimensions, which is exactly calculable but still nontrivial. It would be interesting to explore the properties of the theory further, for instance, by computing non-planar corrections or higher-point functions. Also interesting is to study correlators involving operators outside the loop. We can then start exploring the interplay with the conformal bootstrap\cite{BGLM, Gadde}.

Studying the ladders limit would also be useful for the integrability approach itself. To reproduce the result in the ladders limit using integrability, one needs to resum the mirror particles. To our knowledge, it is the only limit where we can predict exact answers after the resummation\fn{The result at strong coupling also contains information about multi-mirror-particle corrections. However, unlike the ladders limit, it only predicts the leading exponentially dominating contribution.}. By studying this limit further, we may be able to learn how to perform the resummation in the hexagon approach and make connection with powerful methods developed for the spectrum, such as the thermodynamic Bethe ansatz and the quantum spectral curve. 

Furthermore, in this simple set-up, it might be possible to ``derive" the hexagon approach from the Feynman diagrams. In the hexagon approach, the structure constants of the DCO's are given in terms of integration of magnon momenta. On the other hand, the perturbative methods in the ladders limit yields integrals over the positions of ends of propagators. Since both integrals are rather simple to analyze, it may be possible to make direct relation between the two\fn{A naive guess is that they are related by the Fourier transformation. It might also be interesting to explore the relation with the Q-functions since in similar set-ups, it was shown that the Schr\"{o}dinger equation and the Q-functions are related \cite{QSCladder}.}. That would help to clarify the origin of the integrability in $\mathcal{N}=4$ SYM. 
\subsubsection*{AdS dual of the ladders limit}
Yet another interesting problem would be to understand the AdS dual of the ladders limit. A general straight line Wilson loop is dual to a probe worldsheet living in the $AdS_2$ subspace of $AdS_5\times S^5$. In the ladders limit, the worldsheet becomes tensionless while its boundary term remains nontrivial. Although the tensionless limit is in general difficult to analyze, it might be possible to make some progress in this case since one knows the exact answer after the resummation of diagrams. It would be interesting to try to write down the worldsheet action which reproduces the solutions to the Schwinger-Dyson equation.

Recently, certain double-scaling limits of $\mathcal{N}=4$ SYM and related theories are advocated as the simplest examples of the integrable gauge/string duality \cite{stronglydeformed1,stronglydeformed2,stronglydeformed3,stronglydeformed4,stronglydeformed5,stronglydeformed6}. These theories share some features with the ladders limit: First, they have a tensionless limit of the string theory as the bulk dual. Second, in the integrability description, they both arise in the limit where the coupling constant vanishes and the twist diverges. In this sense, the ladders limit may serve as a toy model for such theories. Hopefully understanding the worldsheet description of the ladders limit may be used as a first step towards constructing the bulk dual of such theories.

Lastly, we note that similar resummation of the diagrams was performed recently in \cite{SYK,SYK2} to determine the cubic coupling of the bulk dual of the SYK model. Although the SYK model is physically quite different from the Wilson loop defect CFT since its bulk dual contains gravity, it might still be interesting to ask if the analysis of the ladders limit has any bearing on this interesting new class of holography.

\subsection*{Acknowledgement}
We acknowledge helpful discussions with Z.~Bajnok, D.~Correa and N.~Drukker. S.K. thanks P.~Vieira for a useful conversation. 
The research of M.K. was supported by a postdoctoral fellowship of the Hungarian Academy of Sciences, a Lend\"ulet grant, OTKA 116505 and by the South African Research Chairs Initiative of the Department of Science and Technology and National Research
Foundation, whereas the researches of N.K. and T.N. are supported in part by JSPS Research Fellowship for Young Scientists, from the Japan Ministry of Education, Culture, Sports, Science and Technology. The research of S.K. was supported in part by Perimeter Institute for Theoretical Physics. Research at Perimeter Institute is supported by the Government of Canada through the Department of Innovation, Science and Economic Development Canada and by the Province of Ontario through the Ministry of Research, Innovation and Science.

\appendix

\section{Basic integrals}\label{Y123}
Here we introduce basic integrals following \cite{BKPSS}, which appear in the perturbative computation. The first integral is the so-called 1-loop conformal integral which is defined by
\beq
X_{1234}\equiv \int  \frac{d^{4}x_5}{x_{15}^2x_{25}^2x_{35}^2x_{45}^2}\comma
\eeq
with $x_{ij}= |x_i-x_j|$.
This integral can be evaluated explicitly \cite{UD} as
\beq
X_{1234}=\frac{\pi^2 \Phi(z,\bar{z})}{x_{13}^2x_{24}^2}\comma  \qquad \Phi (z,\bar{z})\equiv \frac{2 {\rm Li}_2 (z)-2 {\rm Li}_2 (\bar{z})+\log z\bar{z}\log \frac{1-z}{1-\bar{z}}}{z-\bar{z}}\comma
\eeq
where $z$ and $\bar{z}$ are the usual conformal cross ratios:
\beq
z\bar{z}=\frac{x_{12}^2x_{34}^2}{x_{13}^2x_{24}^2}\comma \qquad \qquad (1-z)(1-\bar{z})=\frac{x_{14}^2x_{23}^2}{x_{13}^2x_{24}^2}\period
\eeq

Another integral which shows up often is the three-point integral given by
\begin{align}
Y_{123}&\equiv \int \frac{d^{4}x_5}{x_{15}^{2}x_{25}^{2}x_{35}^{2}}\period
\end{align}
It can be evaluated using $X_{1234}$ as
\beq
\begin{aligned}
Y_{123}&= \lim_{x_4 \to \infty}x_4^2 X_{1234}=\frac{\pi^2\Phi (z^{\prime},\bar{z}^{\prime})}{x_{13}^2}\comma 
\end{aligned}
\eeq
with
\beq
z^{\prime}\bar{z}^{\prime}=\frac{x_{12}^2}{x_{13}^2}\comma \qquad \qquad (1-z^{\prime})(1-\bar{z}^{\prime})=\frac{x_{23}^2}{x_{13}^2}\period
\eeq

When all external points are on a single line, these integrals further simplify to
\beq
\begin{aligned}
X_{1234}\big|_{\text{line}}&=-2\pi ^2 \left(\frac{\log (|\tau_{12}\tau_{34}|)}{\tau_{13}\tau_{24}\tau_{14}\tau_{23}}+\frac{\log (|\tau_{13}\tau_{24}|)}{\tau_{12}\tau_{34}\tau_{14}\tau_{32}}+\frac{\log (|\tau_{14}\tau_{23}|)}{\tau_{12}\tau_{43}\tau_{13}\tau_{42}}\right)\comma\\
 Y_{123}\big|_{\text{line}}&=-2\pi ^2 \left(\frac{\log |\tau_{12}|}{\tau_{13}\tau_{23}}+\frac{\log |\tau_{13}|}{\tau_{12}\tau_{32}}+\frac{\log |\tau_{23}|}{\tau_{21}\tau_{31}}\right)\comma
\end{aligned}
\eeq
where $\tau_i$'s are the positions of the external points on the line and $\tau_{ij}=\tau_i-\tau_j$.


\section{Vertex and self-energy diagrams for the three-point functions}\label{ap:B}
In this appendix, we present the detail of the computation of vertex and self-energy diagrams that appear in the three-point functions of DCO's at two loops. As explained in the main text we focus on the terms proportional to $(n_{12}\cdot n_{23}-1)$.

The contributions from the diagrams listed in figure \ref{3ptvertex} can be determined in a way similar to the ones in section \ref{subsec:2pt2loop}. Essentially the only difference is the range of the integration. By changing the range of the integration of the analogous diagrams in section \ref{subsec:2pt2loop}, one obtains
\beq
\begin{aligned}
{\tt T1}=& -\frac{\lambda^2 (n_{12}\cdot n_{23}-1)}{4(4\pi^2)^3}\int_{t_1^{+}}^{t_2^{-}}d\tau_1\int_{t_1^{+}}^{t_2^{-}}d\tau_2\int_{t_2^{+}}^{t_3^{-}}d\tau_3 \,\epsilon (\tau_1-\tau_2) \del_{\tau_1} Y_{123}\comma\\
{\tt T2}=&-\frac{\lambda^2 (n_{12}\cdot n_{23}-1)}{4(4\pi^2)^3}\int_{t_1^{+}}^{t_2^{-}}d\tau_{1}\int_{t_{2}^{+}}^{t_{3}^{-}}d\tau_{2}\int_{t_{2}^{+}}^{t_{3}^{-}}d\tau_{3}\,\epsilon(\tau_{2}-\tau_{3})\del_{\tau_2}Y_{123}\comma\\
{\tt T3}=&-\frac{\lambda^2 (n_{12}\cdot n_{23}-1)}{4(4\pi^2)^3}\int_{-\infty}^{t_{1}^{-}}d\tau_{1}\int_{t_{1}^{+}}^{t_{2}^{-}}d\tau_{2}\int_{t_{2}^{+}}^{t_3^{-}}d\tau_{3}\left(\del_{\tau_2}Y_{123}-\del_{\tau_3}Y_{123}\right)\comma
\\
{\tt T4}=&-\frac{\lambda^2 (n_{12}\cdot n_{23}-1)}{4(4\pi^2)^3}\int_{t_1^{+}}^{t_{2}^{-}}d\tau_{1}\int_{t_{2}^{+}}^{t_{3}^{-}}d\tau_{2}\int_{t_{3}^{+}}^{\infty}d\tau_{3}\left(\del_{\tau_1}Y_{123}-\del_{\tau_2}Y_{123}\right)\period
\end{aligned}
\eeq
As in the computation for the two-point function, we then perform the integration by parts and decompose the integrals into the $\delta$-function terms and the boundary contributions. Here again, the $\delta$-function terms are cancelled by the self-energy diagrams and what remains is
\beq
\begin{aligned}
\left.{\tt T}\right|_{\rm finite}=&-\frac{\lambda^2 (n_{12}\cdot n_{23}-1)}{4(4\pi^2)^3}\left[\int_{t_1^{+}}^{t_2^{-}}d\tau_2\int_{t_2^{+}}^{t_3^{-}}d\tau_3 \left(Y_{t_2^{-}23}+Y_{t_1^{+}23}+Y_{t_3^{-}23}+Y_{t_2^{+}23}\right)\right.\\
&+\int_{-\infty}^{t_1^{-}}d\tau_2\int_{t_2^{+}}^{t_3^{-}}d\tau_3 (Y_{t_2^{-}23}-Y_{t_1^{+}23})+\int_{-\infty}^{t_1^{-}}d\tau_2\int_{t_1^{+}}^{t_2^{-}}d\tau_3(Y_{t_2^{+}23}-Y_{t_3^{-}23})\\
&\left.+\int_{t_2^{+}}^{t_3^{-}}d\tau_2\int_{t_3^{+}}^{\infty}d\tau_3 (Y_{t_2^{-}23}-Y_{t_1^{+}23})+\int_{t_1^{+}}^{t_2^{-}}d\tau_2\int_{t_3^{+}}^{\infty}d\tau_3 (Y_{t_2^{+}23}-Y_{t_3^{-}23})\right]\period
\end{aligned}
\eeq 
One can then perform the integral to get
\begin{align}
\left.{\tt T}\right|_{\rm finite}=-(n_{12}\cdot n_{23}-1)\left(\frac{\lambda}{8\pi^{2}}\right)^2\left(3\zeta(3)+\frac{\pi^{2}}{3}\log\left|\frac{t_{12}t_{23}}{t_{31}\epsilon}\right|\right)\period
\end{align}
\section{Excited states and conformal descendants}\label{ap:sch}
In this appendix, we explain the relation between the vertex function $\Gamma_{\epsilon}$ and the Schr\"{o}dinger equation in \cite{ESSZ,CHMS}. In particular, we clarify the physical meaning of the wave functions of the Schr\"{o}dinger equation by showing that they correspond to the three-point functions of DCOs, and that the excited states of the Schr\"{o}dinger equation correspond to conformal descendants.

For this purpose, let us quickly review how the Schr\"{o}dinger equation comes about from the differential equation for $\Gamma_{\epsilon}$ \eqref{eq:diff_eq2}. To begin with, we rewrite the equation in terms of the ``radial coordinate"\fn{If we instead rewrite the equation in terms of the coordinates $s=S+T$ and $t=S-T$, one arrives at the ``conformal quantum mechanics" \cite{CQM}; the Schr\"{o}dinger equation with the inverse square potential. This description, however, is not very useful for our purpose and we will not discuss it here.}
\beq
S=\exp (-x+y)\comma\qquad T=\exp (x+y)\comma
\eeq
to get
\beq\label{differentialinxy}
\left[-\frac{1}{4}(\del_x^2-\del_y^2)-\frac{\hat{\lambda}}{4\pi^2}\frac{1}{(2\cosh x)^2}\right]\Gamma_{\epsilon}=0\period
\eeq
Physically this rewriting corresponds to considering the theory on $R\times S^{3}$: $x$ describes the (Euclidean) time difference of the two endpoints while $y$ corresponds to the time of the ``center of mass". Then, assuming the form of the solution to be 
\beq\label{gammaexpand}
\Gamma_{\epsilon}=\sum_N c_N e^{\Omega_N y}\Psi_N (x)\comma
\eeq
one can reduce the differential equation \eqref{differentialinxy} to the following one-dimensional Schr\"{o}dinger equation:
\begin{align}
\left[ -\frac{d^2}{dx^2} -\frac{\hat{\lambda}}{4\pi^2} \frac{1}{\cosh^2 x} \right] \Psi_N(x) = -\Omega^2_N\Psi_N(x) \label{eq:schrodinger} \period
\end{align}
The Schr\"{o}dinger equation with this potential (called the P\"{o}schl-Teller potential) has the SL(2,$\mathbb{R}$) symmetry\fn{The difference from the usual conformal quantum mechanics \cite{CQM} lies in that the ``dilatation generator" of the SL(2,$\mathbb{R}$) is identified not with the Hamiltonian itself but with its square root. See for instance \cite{PTsl2}.} and is known to be exactly solvable. This can be seen explicitly by the change of the variable
\beq
z=\frac{1}{1+e^{x}}\comma
\eeq
which maps the problem to the hypergeometric differential equation. 

By using the explicit form of $\Gamma_{\epsilon}$ shown in \eqref{defgammepsi}, one can determine which wave functions appear in the expansion \eqref{gammaexpand}. The result turns out to be given by a sum of two families of solutions\fn{As in the main text $\Omega$ is defined by 
\beq
\Omega (\hat{\lambda})=\frac{1}{2}\left[-1+\sqrt{1+\frac{\hat{\lambda}}{\pi^2}}\right]\period
\eeq
}
\beq\label{gamexpandphys}
\Gamma_{\epsilon}=\sum_{n=0}^{\infty} c_n e^{\Omega^{(n)} y}\Psi_n (x)+\sum_{n=0}^{\infty} \tilde{c}_n e^{\tilde{\Omega}^{(n)} y}\tilde{\Psi}_n (x)
\eeq
with
\beq
\begin{aligned}\label{wavepsipsi}
&\Omega^{(n)}=\Omega-n \comma\qquad \qquad \tilde{\Omega}^{(n)}=-\Omega-n-1 \comma\\
&\Psi_n (z)=\left(z (1-z)\right)^{\frac{\Omega^{(n)}}{2}}{}_{2}F_{1}(\Omega^{(n)}-\Omega,\Omega^{(n)}+\Omega+1,1+\Omega^{(n)};z)\comma\\
&\tilde{\Psi}_n (z)=\left(z (1-z)\right)^{\frac{\tilde{\Omega}^{(n)}}{2}}{}_{2}F_{1}(\tilde{\Omega}^{(n)}-\Omega,\tilde{\Omega}^{(n)}+\Omega+1,1+\tilde{\Omega}^{(n)};z)\period
\end{aligned}
\eeq
These solutions have several interesting properties. First, they are the only solutions to \eqref{eq:schrodinger} for which the hypergeometric function reduces to a polynomial. Second, the first family of solutions with $n< \Omega$ decay at $x=\pm \infty$ and correspond to the bound states of the Schr\"{o}dinger equation \eqref{eq:schrodinger}, as discussed in \cite{CHMS}. Note also that, to reconstruct $\Gamma_{\epsilon}$, one needs to include ``unphysical solutions" which blow up at $x=\pm \infty$, in addition to such bound state solutions. Although it might seem counter-intuitive, it has natural interpretation in terms of the OPE in the defect CFT as we see below.

To see this, recall that the vertex function is obtained as a limit of the four-point ladder kernel $\Gamma_{\epsilon} (S,T)\equiv K(-S,-\epsilon/2\,\,|\,\,\epsilon/2,T)$. A crucial observation is that the ladder kernel itself can be interpreted as a certain four-point function of (trivial) DCOs,
\beq\label{vertexisfour}
\Gamma_{\epsilon}(S,T)=K(-S,-\epsilon/2\,\,|\,\,\epsilon/2,T)=\langle\mathcal{O}^{\circ}_1 (-S)\mathcal{O}^{\circ}_2 (-\epsilon/2)\mathcal{O}^{\circ}_3 (\epsilon/2)\mathcal{O}^{\circ}_{4}(T) \rangle\comma
\eeq
 and the limit $\epsilon \to 0$ corresponds to the OPE limit where $\mathcal{O}_2$ and $\mathcal{O}_3$ approach. Using the OPE, one can replace the product of $\mathcal{O}_2$ and $\mathcal{O}_3$ with an infinite sum\fn{Here we used the fact that the trivial DCOs have zero conformal dimensions.},
\beq\label{infsum23}
\mathcal{O}^{\circ}_2 (-\epsilon/2)\mathcal{O}^{\circ}_3 (\epsilon/2) = \sum_{\tilde{\mathcal{O}}} \epsilon^{\Delta_{\tilde{\mathcal{}O}}}c_{23 \tilde{\mathcal{O}}}\tilde{\mathcal{O}}(0)\period
\eeq
Here the sum on the right hand side is over both primaries and descendants, and $c_{23\mathcal{\hat{\mathcal{O}}}}$ denotes the structure constant. Using this OPE inside the four-point function \eqref{vertexisfour}, we get the following infinite-sum representation for the vertex function
\beq
\begin{aligned}
\Gamma_{\epsilon}(S,T)&=\sum_{\tilde{\mathcal{O}}} \epsilon^{\Delta_{\tilde{\mathcal{}O}}}c_{23 \tilde{\mathcal{O}}}\langle \mathcal{O}^{\circ}_1 (-S)\tilde{\mathcal{O}}(0)\mathcal{O}^{\circ}_{4}(T)\rangle\period
\end{aligned}
\eeq
Let us now compare this sum with the sum over wave functions \eqref{gamexpandphys}. To do so, one has to know the behavior of $\langle \mathcal{O}^{\circ}_1 (-S)\tilde{\mathcal{O}}(0)\mathcal{O}^{\circ}_{4}(T)\rangle$ (both for primaries and descendants) and express it in terms of the $x$ and $y$ coordinates. When $\tilde{\mathcal{O}}$ is primary, the behavior of the three-point function is well-known\fn{For the sake of brevity, below we omit writing the subscript $\tilde{O}$ in $\Delta_{\tilde{O}}$.}, 
\beq\label{prim3pt}
\langle \mathcal{O}^{\circ}_1 (-S)\tilde{\mathcal{O}}_{\rm primary}(0)\mathcal{O}^{\circ}_{4}(T)\rangle\propto \left(\frac{S+T}{ST}\right)^{\Delta}\period
\eeq
On the other hand, the behavior for the descendants can be computed by differentiation as
\beq\label{eq120}
\langle \mathcal{O}^{\circ}_1 (-S)\del^{n}\tilde{\mathcal{O}}_{\rm primary}(0)\mathcal{O}^{\circ}_{4}(T)\rangle\propto \left(\frac{S+T}{ST}\right)^{\Delta}\sum_{k=0}^{n}\pmatrix{c}{n\\k}(-1)^{k}(\Delta)_k (\Delta)_{n-k}\frac{1}{S^{k}T^{n-k}}\comma
\eeq
with $(x)_k$ being the Pochhammer symbol. Re-expressing this in terms of $x$ and $y$, we obtain
\beq
\begin{aligned}\label{toconvertto}
\eqref{eq120}&= e^{-(\Delta+n)y}(e^{x}+e^{-x})^{\Delta+n}\sum_{k=0}^{n}\pmatrix{c}{n\\k}(-1)^{k}(\Delta)_k (\Delta)_{n-k}e^{2 k x}(1+e^{-2 x})^{-n}\\
&=e^{-(\Delta+n)y} (z (1-z))^{-\frac{\Delta+n}{2}}\left[\sum_{k=0}^{n}\pmatrix{c}{n\\k}(-1)^{k}(\Delta)_k (\Delta)_{n-k}z^{n-k}(1-z)^{k}\right]\period
\end{aligned}
\eeq
In the second line, we further rewrote it in terms of $z=1/(1+e^{x})$. The polynomial in the bracket turns out to be summed into a hypergeometric function ${}_2 F_1 (-n,1-2\Delta-n,1-\Delta-n,z)$. We thus get the expression
\beq
\begin{split}
&\langle \mathcal{O}^{\circ}_1 (-S)\del^{n}\tilde{\mathcal{O}}_{\rm primary}(0)\mathcal{O}^{\circ}_{4}(T)\rangle\\&\propto e^{-(\Delta+n)y} (z (1-z))^{-\frac{\Delta+n}{2}} {}_2 F_1 (-n,1-2\Delta-n,1-\Delta-n,z)\period
\end{split}
\eeq

With the identifications $\Delta=-\Omega$ and $\Delta=1+\Omega$, this coincides with $e^{\Omega_n y}\Psi_{n}$ and $e^{\tilde{\Omega}_n y}\tilde{\Psi}_{n}$ in \eqref{wavepsipsi} respectively. We can therefore interpret the sum \eqref{gamexpandphys} really as the OPE expansion and the wave functions are identified with the three-point functions:
\beq
\begin{aligned}
&\Gamma_{\epsilon}=\sum_{X={\rm DCO}, {\rm shadow}}\sum_{n=0}^{\infty}\epsilon^{\Delta_X +n}c_{X,n} \langle \mathcal{O}^{\circ}_1 (-S)\del^{n}\mathcal{O}^{\bullet}_{X}(0)\mathcal{O}^{\circ}_{4}(T)\rangle\comma\\
&\langle \mathcal{O}^{\circ}_1 (-S)\del^{n}\mathcal{O}^{\bullet}_{\rm DCO}(0)\mathcal{O}^{\circ}_{4}(T)\rangle\quad  \leftrightarrow \quad e^{\Omega_n y}\Psi_{n}\comma\\
&\langle \mathcal{O}^{\circ}_1 (-S)\del^{n}\mathcal{O}^{\bullet}_{\rm shadow}(0)\mathcal{O}^{\circ}_{4}(T)\rangle \quad \leftrightarrow \quad e^{\tilde{\Omega}_n y}\tilde{\Psi}_{n}\period
\end{aligned}
\eeq
Here $\mathcal{O}^{\bullet}_{\rm DCO}$ is a nontrivial DCO, which we studied in the main text, and $\mathcal{O}^{\bullet}_{\rm shadow}$ is its {\it shadow} operator\fn{In unitary CFTs, the shadow operators do not usually show up in the spectrum since they are often below the unitarity bound. However, the possibility of having both an operator and its shadow in the spectrum is not totally ruled out. In fact, it is known that some long-range CFTs have such a spectrum.}, which has dimension $\Delta_{\rm shadow}=1-\Delta_{\rm DCO}=1+\Omega$. 
This provides a clear physical interpretation of the wave functions for the Schr\"{o}dinger equation \eqref{eq:schrodinger}.
\section{Contribution from the integral of $\Gamma_{\rm UV}$}\label{ap:gamUV}
In this appendix, we show that, in the $\epsilon\to 0$ limit, the integrals involving the vertex function $\Gamma_{\epsilon}$ can be approximated by replacing $\Gamma_{\epsilon}$ with its IR counterpart, $\Gamma_{\rm IR}$. More precisely the goal is to show that the ratio between the contributions from $\Gamma_{\rm UV}$ and $\Gamma_{\rm IR}$ is given as follows:
\begin{align}
\frac{\int ds \int dt \,\Gamma_{\rm UV}(s,t) f(s,t)}{\int ds \int dt \,\Gamma_{\rm IR}(s,t) f(s,t)} \leq O\left(\epsilon\log \epsilon\right)\, \overset{\epsilon\to 0}{\to }0 \period \label{eq:IR_integral}
\end{align}
Here $f(s,t)$ denotes the rest of the integrand, which may contain other vertex functions, propagators and the ladder kernels $K$.

For this purpose, it is convenient to split the vertex function in a slightly different way as follows:
\beq
\begin{aligned}\label{gamtiluv}
&\Gamma_{\epsilon}(u)=\tilde{\Gamma}_{\rm IR}(u)+\tilde{\Gamma}_{\rm UV}(u)\comma\\
&\tilde{\Gamma}_{\rm IR}(u)=\frac{A(\Omega)}{(1-u)^{\Omega}}\period
\end{aligned}
\eeq
Since the ratio $(\Gamma_{\rm IR}-\tilde{\Gamma}_{\rm IR})/\Gamma_{\rm IR}$ is always of order $O(\epsilon)$ (regardless of their arguments), it is enough to show \eqref{eq:IR_integral} for $\tilde{\Gamma}_{\rm IR}$ and $\tilde{\Gamma}_{\rm UV}$.

Now, let us estimate the maximal value of $\tilde{\Gamma}_{\rm UV}$. In all the examples studied in the main text, the cross ratio $u=\frac{(S-\epsilon/2)(T-\epsilon/2)}{(S+\epsilon/2)(T+\epsilon/2)}$ takes values in $[0,1-\epsilon/C]$\fn{$u$ can reach $1$ only when $S=T=\infty$. However, we never encounter an integral whose integration regions both extend to infinity.} with a $O(1)$ positive constant $C$. In this region, the UV vertex $\tilde{\Gamma}_{\rm UV}$ monotonically decreases in $u$ for $\Omega\leq 1$ while it monotonically increases in $u$ for $\Omega>1$\fn{One can easily verify this by using the definitions of the vertex functions \eqref{defgammepsi} and \eqref{gamtiluv}, and the series expansion of the hypergeometric function.}. Therefore, the maximal absolute value of the UV vertex is given by
\beq
{\rm max}\,|\tilde{\Gamma}_{\rm UV}(u)|=\left\{ \begin{array}{llc} |1-A(\Omega)|&(=|\tilde{\Gamma}_{\rm UV}(0)|) & {\rm for} \ \Omega \le 1 \\
O\left( \epsilon^{1-\Omega} \right)&(=|\tilde{\Gamma}_{\rm UV}(1-\epsilon/C)|)  &  {\rm for} \ \Omega>1 \end{array} \period\right.
\eeq
Hence, the integral of $\tilde{\Gamma}_{\rm UV}$ can be bounded from above as follows:
\beq
\int ds \int dt \,\tilde{\Gamma}_{\rm UV}(s,t) f(s,t)\leq \frac{\tilde{C}}{\epsilon^{\Omega-1}} \int ds \int dt \, f(s,t)\period
\eeq
In all the cases encountered in the main text, the integral of $f(s,t)$ can produce at most logarithmic divergences\fn{This inverse square behavior comes from a propagator contained in $f (s,t)$.} $\int ds \int dt (s-t)^{-2}\sim \log \epsilon$. We thus have
\beq\label{estimation1}
\int ds \int dt \,\tilde{\Gamma}_{\rm UV}(s,t) f(s,t)\leq O\left(\frac{\log \epsilon}{\epsilon^{\Omega-1+|f|}}\right)\comma
\eeq
where $\epsilon^{-|f|}$ is the singularity contained already in the integrand, $f \sim O (\epsilon^{-|f|})$.

On the other hand, since $\tilde{\Gamma}_{\rm IR}\sim \epsilon^{-\Omega} \times k(s,t)$ with $k(s,t)$ being the $O(1)$ function, we can easily estimate its integral as
\beq\label{estimation2}
\int ds \int dt \tilde{\Gamma}_{\rm IR} (s,t) f(s,t)\geq O\left(\frac{1}{\epsilon^{\Omega+|f|}}\right)\period
\eeq
Combining \eqref{estimation1} and \eqref{estimation2}, we get the estimation \eqref{eq:IR_integral} for $\tilde{\Gamma}_{\rm UV}$ and $\tilde{\Gamma}_{\rm IR}$.
\section{Evaluation of the integral \eqref{simpleonenontrivial}}\label{ap:integral}
Here we compute the integral
\begin{align}
\begin{split}
{\tt integral}&\equiv \int^{\tau_2}_{\tau_1}ds\, \del_s \Gamma^{R}(\infty,s-\tau_1)\, K(s,\tau_2|\tau_3,\infty)\\
&=\frac{\sqrt{A(\Omega)}}{\tau_{32}^{\Omega}}\int^{\tau_2}_{\tau_1} ds\,\del_s \left[(s-\tau_1)^{\Omega}\right](\tau_3-s)^{\Omega}\,\,{}_{2}F_{1}\left(-\Omega,-\Omega,1;\frac{\tau_2-s}{\tau_3-s}\right)\period
\end{split}
\end{align}
As a first step, we perform the change of variables,
\beq
x=\frac{s-\tau_2}{s-\tau_3}\frac{\tau_1-\tau_3}{\tau_1-\tau_2}\comma
\eeq
and use the identity ${}_2F_{1}(a,b,c;z)=(1-z)^{-a-b+c}{}_2F_{1}(c-a,c-b,c;z)$ to get
\beq
{\tt integral}=\Omega\sqrt{A(\Omega)}\tau_{21}^{\Omega} (1-\alpha)\int^{1}_{0} dx\,(1- x)^{\Omega-1}\,\,{}_{2}F_{1}\left(\Omega+1,\Omega+1,1;\alpha x\right)\comma
\eeq
with $\alpha=\tau_{21}/\tau_{32}$. To proceed, we rewrite ${}_2F_{1}$ using the integral representation as
\beq
{}_{2}F_{1}\left(\Omega+1,\Omega+1,1;\alpha x\right)=\frac{1}{\Gamma(\Omega+1)\Gamma(-\Omega)}\int^{1}_{0}dy\, y^{\Omega}(1-y)^{-\Omega-1} (1-y\alpha x)^{-\Omega-1}\period
\eeq
One can then perform the $x$ integral to get\fn{Here we used the integral expression for the hypergeometric function and the identity ${}_{2}F_{1}(a,1,a;z)=(1-z)^{-1}$.}
\beq
{\tt integral}=\frac{\sqrt{A(\Omega)}\tau_{21}^{\Omega} (1-\alpha)}{\Gamma(\Omega+1)\Gamma(-\Omega)}\int^{1}_{0} dy\,y^{\Omega}(1-y)^{-\Omega-1}(1-\alpha y)^{-1}\period
\eeq
This is again a hypergeometric integral and we can compute it as follows:
\beq
{\tt integral}=\sqrt{A(\Omega)}\tau_{21}^{\Omega} (1-\alpha){}_{2}F_{1}\left(1,\Omega+1,1;\alpha \right)\period
\eeq
Finally, using the identity ${}_{2}F_{1}\left(1,\Omega+1,1;\alpha \right)=(1-\alpha)^{-\Omega-1}$, we arrive at
\beq
{\tt integral}=\frac{\sqrt{A(\Omega)}}{\tau_{21}^{-\Omega}\tau_{32}^{\Omega}\tau_{31}^{-\Omega}}\period
\eeq
\section{An infinite sum representation for $C_{\bullet\bullet\bullet}$}\label{ap:sum}
Here we explicitly evaluate the integral representation for $C_{\bullet\bullet\bullet}$ \eqref{case3integral}, and derive an infinite-sum representation. First, we perform the following change of variables:
\beq
x=\frac{1-\bar{u}}{1-(1-\bar{s})\bar{u}}\comma\quad y=1-(1-\bar{s})\bar{u} \comma\quad z=\frac{1-\bar{t}}{1-(1-\bar{u})\bar{t}}\period
\eeq
After taking into account the Jacobian of the transformation, $\frac{y}{(1-xyz)^2}$, we get
\begin{align}
\begin{split}
J &=\prod_{i=1}^3 \Omega_i \int_0^1 dx \int_0^1 dy \int_0^1 dz x^{\Omega_1-1}y^{\Omega_1} z^{\Omega_3-1} \\
&(1-x)^{\Omega_1+1} (1-y)^{\Omega_2-1}(1-z)^{\Omega_2+1} (1-xy)^{-(\Omega_1+\Omega_2-\Omega_3+1)}(1-yz)^{-(\Omega_2+1)} \period
\end{split} \label{eq:int_xyz}
\end{align}
The integrals of $x$ and $z$ yield the hypergeometric functions,
\begin{align}
\begin{split}
J=&\prod_{i=1}^3 \Omega_i  \int_0^1 dy  y^{\Omega_1} (1-y)^{\Omega_2-1} \frac{\Gamma(\Omega_3) \Gamma(\Omega_2+2)}{\Gamma(\Omega_2+\Omega_3+2)} {}_2F_1(\Omega_3,\Omega_2+1,\Omega_2+\Omega_3+2;y)   \\
&\times  \frac{\Gamma(\Omega_1) \Gamma(\Omega_1+2)}{\Gamma(2\Omega_1+2)} {}_2F_1(\Omega_1,\Omega_1+\Omega_2-\Omega_3+1,2\Omega_1+2;y)   \period
\end{split}
\end{align}
The remaining $y$ integral can be performed using the series expansion of the hypergeometric function and the Euler integral representation for the generalized hypergeometric function:
\begin{align}
\begin{split}
{}_3F_2 &\left( \begin{matrix} \alpha_1,\alpha_2,\alpha_3 \\ \beta_1 , \beta_2 \end{matrix};z \right) = \frac{\Gamma (\beta_1)\Gamma(\beta_2)}{\Gamma(\alpha_1) \Gamma(\beta_1-\alpha_1) \Gamma(\alpha_2)\Gamma(\beta_2-\alpha_2)} \\
\times & \int_0^1 ds \int_0^1 dt s^{\alpha_1-1}(1-s)^{\beta_1-\alpha_1-1} t^{\alpha_2-1}(1-t)^{\beta_2-\alpha_2-1} (1-zst)^{-\alpha_3} \period
\end{split}
\end{align}
Finally, we obtain the following expression for the structure constant
\begin{align}
\begin{split}
&C_{\bullet \bullet \bullet} = \prod_{k=1}^3 \sqrt{A(\Omega_k)}\frac{\Gamma(\Omega_1+1)^2\Gamma(\Omega_1+2)\Gamma(\Omega_2+1)\Gamma(\Omega_2+2)\Gamma(\Omega_3+1)}{\Gamma(2\Omega_1+2)\Gamma(\Omega_2+\Omega_3+2) \Gamma(\Omega_1+\Omega_2+1)}  \\
&\times \sum_{k=0}^{\infty} \frac{(\Omega_3)_k(\Omega_2+1)_k}{(\Omega_2+\Omega_3+2)_kk!} \frac{(\Omega_1+1)_k}{(\Omega_1+\Omega_2+1)_k} {}_3F_2 \left( \begin{matrix} \Omega_1,\Omega_1+\Omega_2-\Omega_3+1,\Omega_1+k+1 \\ 2\Omega_1+2 , \Omega_1+\Omega_2+k+1 \end{matrix};1 \right) \period \label{eq:str_3_sum}
\end{split}
\end{align}
Unlike the integral representation \eqref{case3integral}, this expression is not manifestly symmetric under the permutation of $\Omega_i$'s. One can nevertheless check easily that the expression correctly reproduces $C_{\bullet\bullet\circ}$ by sending one of $\Omega_i$'s to zero.

\end{document}